\documentclass{emulateapj}
\usepackage{graphicx}
\usepackage{epsfig}
\usepackage{natbib}
\usepackage{multirow}
\newcommand{\km}{{\rm\thinspace km}}

\newcommand{\s}{{\rm\thinspace s}}

\newcommand{\erg}{{\rm\thinspace ergs}}

\newcommand{\ergps}{\hbox{$\erg\s^{-1}\,$}}

\newcommand{\kmps}{\hbox{$\km\s^{-1}\,$}}

\newcommand{\halpha}{H$\alpha$}

\newcommand{\hii}{H\thinspace{\sc ii}}
\newcommand{\nii}{[N\thinspace{\sc ii}]}
\newcommand{\none}{[N\thinspace{\sc i}]}
\newcommand{\mgi}{[Mg\thinspace{\sc i}]}
\newcommand{\oiii}{[O\thinspace{\sc iii}]}
\newcommand{\oii}{[O\thinspace{\sc ii}]}
\newcommand{\oi}{[O\thinspace{\sc i}]}

\newcommand{\neiii}{[Ne\thinspace{\sc iii}]}
\newcommand{\ariv}{[Ar\thinspace{\sc iv}]}
\newcommand{\ariii}{[Ar\thinspace{\sc iii}]}

\newcommand{\feiii}{[Fe\thinspace{\sc iii}]}
\newcommand{\fev}{[Fe\thinspace{\sc v}]}
\newcommand{\nev}{[Ne\thinspace{\sc v}]}
\newcommand{\sii}{[S\thinspace{\sc ii}]}

\newcommand{\heii}{He\thinspace{\sc ii}}
\newcommand{\hei}{He\thinspace{\sc i}}
\newcommand{\siiii}{Si\thinspace{\sc iii}}
\newcommand{\cii}{C\thinspace{\sc ii}}
\newcommand{\ciii}{C\thinspace{\sc iii}}
\newcommand{\civ}{C\thinspace{\sc iv}}
\newcommand{\niii}{N\thinspace{\sc iii}}
\newcommand{\nv}{N\thinspace{\sc v}}

\newcommand{\peaA}{J081552.00+215623.6}
\newcommand{\peaAs}{J081552}
\newcommand{\peaB}{J121903.98+152608.5}
\newcommand{\peaBs}{J121903}
\newcommand{\peaC}{J101157.08+130822.0}
\newcommand{\peaCs}{J101157}
\newcommand{\peaD}{J145735.13+223201.8}
\newcommand{\peaDs}{J145735}
\newcommand{\peaE}{J100918.99+291621.5}
\newcommand{\peaEs}{J100918}
\newcommand{\peaF}{J030321.4-075923}
\newcommand{\peaFs}{J030321}

\newcommand{\pc}{{\rm\thinspace pc}}

\newcommand{\Lsol}{\hbox{\thinspace L$_{\sun}$}}
\newcommand{\Msol}{\hbox{\thinspace M$_{\sun}$}}
\newcommand{\Zsol}{\hbox{\thinspace Z$_{\sun}$}}

\shorttitle{Ionizing Radiation in the Green Peas}
\shortauthors{Jaskot \& Oey}

\begin{document}

\title{The Origin and Optical Depth of Ionizing Radiation in the ``Green Pea'' Galaxies}

\author{A. E. Jaskot and
	M. S. Oey
	}	
\affiliation{University of Michigan, Dept. of Astronomy, 
        830 Dennison Bldg., Ann Arbor, MI 48109, USA.}

\begin{abstract}
Although Lyman continuum (LyC) radiation from star-forming galaxies likely drove the reionization of the Universe, observations of star-forming galaxies at low redshift generally indicate low LyC escape fractions. However, the extreme \oiii/\oii~ratios of the $z=0.1-0.3$ Green Pea galaxies may be due to high escape fractions of ionizing radiation. To analyze the LyC optical depths and ionizing sources of these rare, compact starbursts, we compare nebular photoionization and stellar population models with observed emission lines in the Peas' SDSS spectra. We focus on the six most extreme Green Peas, the galaxies with the highest \oiii/\oii~ratios and the best candidates for escaping ionizing radiation. The Balmer line equivalent widths and \hei~$\lambda$3819 emission in the extreme Peas support young ages of 3-5 Myr, and \heii~$\lambda$4686 emission in five extreme Peas signals the presence of hard ionizing sources. Ionization by active galactic nuclei or high-mass X-ray binaries is inconsistent with the Peas' line ratios and ages. Although stacked spectra reveal no Wolf-Rayet (WR) features, we tentatively detect WR features in the SDSS spectra of three extreme Peas. Based on the Peas' ages and line ratios, we find that WR stars, chemically homogeneous O stars, or shocks could produce the observed \heii~emission. If hot stars are responsible, the Peas' optical depths are ambiguous. However, accounting for emission from shocks lowers the inferred optical depth and suggests that the Peas may be optically thin. The Peas' ages likely optimize the escape of Lyman-continuum radiation; they are old enough for supernovae and stellar winds to reshape the interstellar medium, but young enough to possess large numbers of UV-luminous O or WR stars.

\end{abstract}

\section{Introduction}

Radiative feedback from massive stars has powerful effects on both the interstellar medium (ISM) and intergalactic medium (IGM). In the early Universe, massive stars likely produced most of the ionizing radiation responsible for reionization \citep[e.g.,][]{bolton05,bouwens12,fontanot12}. In addition, high-energy radiation from massive stars is a key component of galactic energy budgets and has important consequences for the ISM. This radiation may suppress or trigger additional star formation \citep{elmegreen77,haiman97} and may escape from \hii~regions to generate the diffuse, warm ionized medium \citep[e.g.,][]{reynolds84,oey97}. 

The fraction of ionizing radiation that escapes \hii~regions and galaxies and the ionizing spectra of massive stars are highly uncertain, however. Near the redshift of reionization, the low-mass galaxies that may contribute most of the ionizing radiation are unobservable \citep{finkelstein12}. Estimates of the galactic escape fraction necessary to sustain reionization range from $< 13\%$ to $\geq 50\%$, depending on the luminosity function and redshift assumed \citep{finkelstein12}. At low redshift, observations of star-forming galaxies give upper limits on the escape fraction of only a few percent \citep[e.g.,][]{leitherer95,heckman01,siana10}. At intermediate redshift, the incidence of detected Lyman continuum radiation is rare \citep[e.g.,][]{steidel01,shapley06} and made uncertain by possible low-redshift contamination \citep{vanzella12}. The intrinsic ionizing spectrum of massive stars is likewise poorly known at all redshifts. This uncertainty translates directly to uncertainties in \hii~region and galactic escape fractions \citep[e.g.,][]{shapley06,voges08}, nebular conditions, and star formation rates. 

Nebular spectra can provide information on both the spectral shape of the incident UV radiation and the optical depth \citep[e.g.,][]{oey00,giammanco05,zastrow11} but these observations are still out of reach for most high redshift galaxies. Instead, many studies have focused on samples of low-redshift galaxies with properties similar to galaxies at high redshift. One such sample is the `Green Pea' galaxies \citep{cardamone09}, a collection of rare emission-line galaxies at $z=0.1-0.3$ identified in the Sloan Digital Sky Survey (SDSS) through their unusually strong \oiii~$\lambda$5007 emission. In particular, due to their extremely high \oiii/\oii~ratios, which may indicate a deficit of low-ionization emission, the Green Peas are an ideal sample to search for optically thin galaxies. The 80 star-forming Peas resemble high-redshift galaxies through their low metallicities, low extinction, high UV luminosities, and enormous specific star formation rates \citep{cardamone09,izotov11,amorin12}. {\it Hubble Space Telescope (HST)} imaging of a handful of the Peas reveals clumpy morphologies and super star clusters \citep{cardamone09}, a common mode of star formation at high redshift \citep[e.g.,][]{cowie95,dickinson03,elmegreen05}. The Green Pea sample therefore represents the best opportunity in the local Universe to study high redshift star-forming conditions and the escape of ionizing radiation.

The precise origin of the Peas' extreme \oiii/\oii~ratios is uncertain. One possibility is that the high ratios are the result of a high ionization parameter, $U$. At a given metallicity, \oiii/\oii~increases with $U$ \citep[e.g.,][]{kewley02,martin-manjon10}. The dimensionless ionization parameter for photoionized gas is defined as
\begin{equation}
\label{u_eqn}
U \equiv  \frac{Q}{4 \pi n_{\rm H} r_0^2 c},
\end{equation}where $Q$ is the hydrogen-ionizing photon emission rate, $n_{\rm H}$ is the total hydrogen density, $r_0$ is the inner nebular radius, and $c$ is the speed of light \citep{osterbrock06}. As powerful starbursts, the Peas likely have compact star forming regions and a hard ionizing radiation field, which may act to increase their ionization parameters and explain their high \oiii/\oii~ratios. Alternatively, \citet{giammanco05} and \citet{pellegrini12} demonstrate that escaping ionizing radiation decreases the emission contribution of low-ionization species. This suppression would also act to increase the \oiii/\oii~ratios and would lead to overestimates of $U$. In this paper, we evaluate the possibility that a low optical depth contributes to the Peas' high observed \oiii/\oii~ratios. We specifically examine the Peas with the highest \oiii/\oii~ratios and consider whether extreme ionization parameters alone are sufficient to explain the observed ratios. 

Several previous works have investigated the ionization conditions in the Peas. In addition to the Peas' powerful \oiii~emission, \citet{hawley12} identifies \heii~$\lambda$4686 emission in nine Peas, providing further evidence for high ionization conditions. The presence of \heii~$\lambda$4686 emission in the Peas may indicate that the Peas are in the middle of a short-lived phase dominated by emission from Wolf-Rayet (WR) stars \citep{hawley12}. Additional support for the WR scenario comes from observations by \citet{amorin12}, who find broad, WR stellar features in deep spectra of three Peas. However, given the small sample size, it is unclear whether these observations are characteristic of the other Peas. 

Studies of the Green Peas and similar galaxies suggest that the Peas may indeed be optically thin. Intermediate redshift Ly$\alpha$ emitters (LAEs) have similar \oiii/\oii~ratios and metallicities to the Green Peas, and fits to the LAEs' spectra imply high escape fractions \citep{nakajima12}. The Green Pea sample overlaps with other nearby samples of galaxies, selected on the basis of high Balmer line equivalent widths \citep{shim12} or high UV surface brightness \citep{heckman05}. Studies of these samples also support a low optical depth. The sample of H$\alpha$~emitters (HAEs) studied by \citet{shim12} resemble WR galaxies in their UV properties and inferred ionization parameters, although the HAEs are younger and have lower metallicities than the WR galaxies. \citet{shim12} propose that the high ionization parameters inferred in the HAEs may instead indicate a low optical depth and a high escape fraction of ionizing photons. This scenario has also been suggested by \citet{overzier09} for a sample of UV-bright Lyman Break Analogs (LBAs), which includes several of the Peas. The LBAs show enhanced \oiii/H$\beta$~ratios, and many of the LBAs have unusually low ratios of H$\alpha$/UV \citep{overzier09}. Although a low optical depth could explain these properties, \citet{overzier09} conclude, that an aged starburst ($>10$ Myr old), dominated by supernova heating, provides a better explanation for the observed spectral properties of the sample. \citet{heckman11} investigated the optically thin scenario by using {\it HST} observations of \cii~$\lambda$1335 to infer the optical depth of several LBAs, including one Green Pea. Although the \cii~line in the Pea is optically thin, \citet{heckman11} point out that the consistency between the galaxy's H$\alpha$, UV, and IR luminosities makes the optically thin hypothesis tenuous. 

Thus, a variety of ionizing sources and optical depth scenarios may explain the ionization conditions in Green Pea galaxies. In this paper, we consider in particular whether a low optical depth could explain the observed line ratios in the Green Peas with the highest \oiii/\oii~ratios. We select six extreme Green Peas from the \citep{cardamone09} sample on the basis of their \oiii/\oii~ratios, as these galaxies are the most likely to be optically thin. We then compare nebular and stellar population models with the observed line ratios in the six most extreme Peas to evaluate the galaxies' optical depths and ionizing sources. We discuss the spectral line measurements and properties of the Green Peas in \S\ref{sec:data}. In \S\ref{sec:analysis}, we analyze the observed line ratios and evaluate the potential ionizing sources. We discuss the results and their consequences for the optical depth of the Peas in \S\ref{sec:discussion} and summarize our results in \S\ref{sec:summary}. We assume a cosmology with $H_0=70$ km s$^{-1}$ Mpc$^{-1}$, $\Omega_m=0.3$, and $\Lambda_0=0.7$ throughout.

\section{Data and Measurements}
\label{sec:data}

The Green Peas are a subset of extreme, compact emission-line galaxies at redshifts between $z=0.1$ and $z=0.3$ \citep{izotov11}. \citet{cardamone09} identify a sample of 103 narrow-line Green Pea galaxies in SDSS Data Release 7 (DR7) through color selection criteria and signal-to-noise (S/N) cuts. The color selection isolates galaxies whose $u-r$, $g-r$, $r-i$, and $r-z$ colors deviate from the normal SDSS galaxy and quasar population due to intense \oiii~$\lambda$5007 emission in the $r$ band \citep{cardamone09}. Using the \citet[BPT]{baldwin81} diagram to separate active galactic nuclei (AGN) and star-forming galaxies, \citet{cardamone09} then identify a sample of 80 starbursting galaxies. Several of these starbursts have high \oiii~$\lambda\lambda$5007,4959/\oii~$\lambda$3727~ratios, implying unusually high ionization parameters \citep[e.g.,][]{martin-manjon10}. The Peas' \oiii/\oii~ratios raise the possibility that the galaxies may be optically thin to Lyman continuum radiation. A low optical depth would suppress the \oii~emission, consistent with the high observed \oiii/\oii~ratios. A less-extreme ionization parameter in combination with a low optical depth could therefore be an alternative explanation for the observed ratios.

We obtained spectra for the 80 starbursting Peas listed in \citet{cardamone09} from the SDSS DR7 \citep{abazajian09}. We then measured the fluxes and equivalent widths (EWs) of the galaxies' emission lines using the IRAF\footnote{IRAF is distributed by the National Optical Astronomy Observatories, which are operated by the Association of Universities for Research in Astronomy, Inc., under cooperative agreement with the National Science Foundation.} task {\tt splot}. As discussed by \citet{amorin12b}, the Peas' emission line profiles have broad wings, indicating the presence of high velocity gas. When necessary, we deblended the H$\alpha$ and \nii~lines using a Voigt profile fit, as a Gaussian fit did not capture the emission line wings. The \sii~$\lambda\lambda$6716 and 6731 lines were deblended with a Gaussian fit, if necessary. We calculated errors in the line measurements according to
\begin{equation}
\label{error_eqn}
\sigma_{\rm line} = \sigma_{\rm rms} \sqrt{N},
\end{equation}where $\sigma_{\rm rms}$~is the average root-mean-square value per pixel measured at either side of the spectral line and $N$ is the number of pixels in the line. 

We simultaneously fit the measured ratios of H$\alpha$/H$\beta$, H$\gamma$/H$\beta$, and H$\delta$/H$\beta$~for both reddening and stellar absorption. We adopted intrinsic ratios of 2.86, 0.468, and 0.259, respectively \citep{storey95}, which are appropriate for $n_{\rm H}=100$ cm$^{-3}$ and $T_e$=10,000 K.  We then corrected the line fluxes for extinction using the \citet{calzetti00} reddening law. In many of the Peas, the \oiii~$\lambda\lambda$5007,4959/$\lambda$4363 ratio indicates an electron temperature much higher than $T_e$=10,000 K. As a result, we re-fit the reddening and stellar absorption for these objects, using the Balmer line ratios at $T_e$=12,500 K or $T_e$=15,000 K from \citet{storey95}. To estimate the appropriate temperature, we used the IRAF {\tt temden} routine, assuming a density of 100 cm$^{-3}$. We used the errors in the H$\alpha$/H$\beta$~ratios to estimate the uncertainties in the extinction correction. We further assumed an uncertainty of 0.04 in the intrinsic H$\alpha$/H$\beta$~ratio, corresponding to the change in the intrinsic ratio when $T_e$ is varied by 2500 K. The intrinsic ratio is less sensitive to density; increasing the density by an order of magnitude changes the intrinsic H$\alpha$/H$\beta$~ratio by only 0.01 or less, depending on the assumed temperature. We then added the reddening uncertainty in quadrature with the flux errors. As noted by \citet{cardamone09}, the Peas generally have low extinction, and our average extinction coefficient $c$(H$\beta$)$=0.27$. The stellar absorption values are also typically low, with a median absorption of 1.05 \AA, indicating that the spectra are dominated by a young population. 

While the Peas generally have high \oiii/\oii~ratios, two Peas (J133940.71+552740.0 and J094347.22+262042.5) actually have stronger \oii~emission than \oiii. Since we are investigating the causes of high \oiii/\oii~ratios in the Peas, we have excluded these objects from further analysis.

\begin{figure*}
\epsscale{1.0}
\plotone{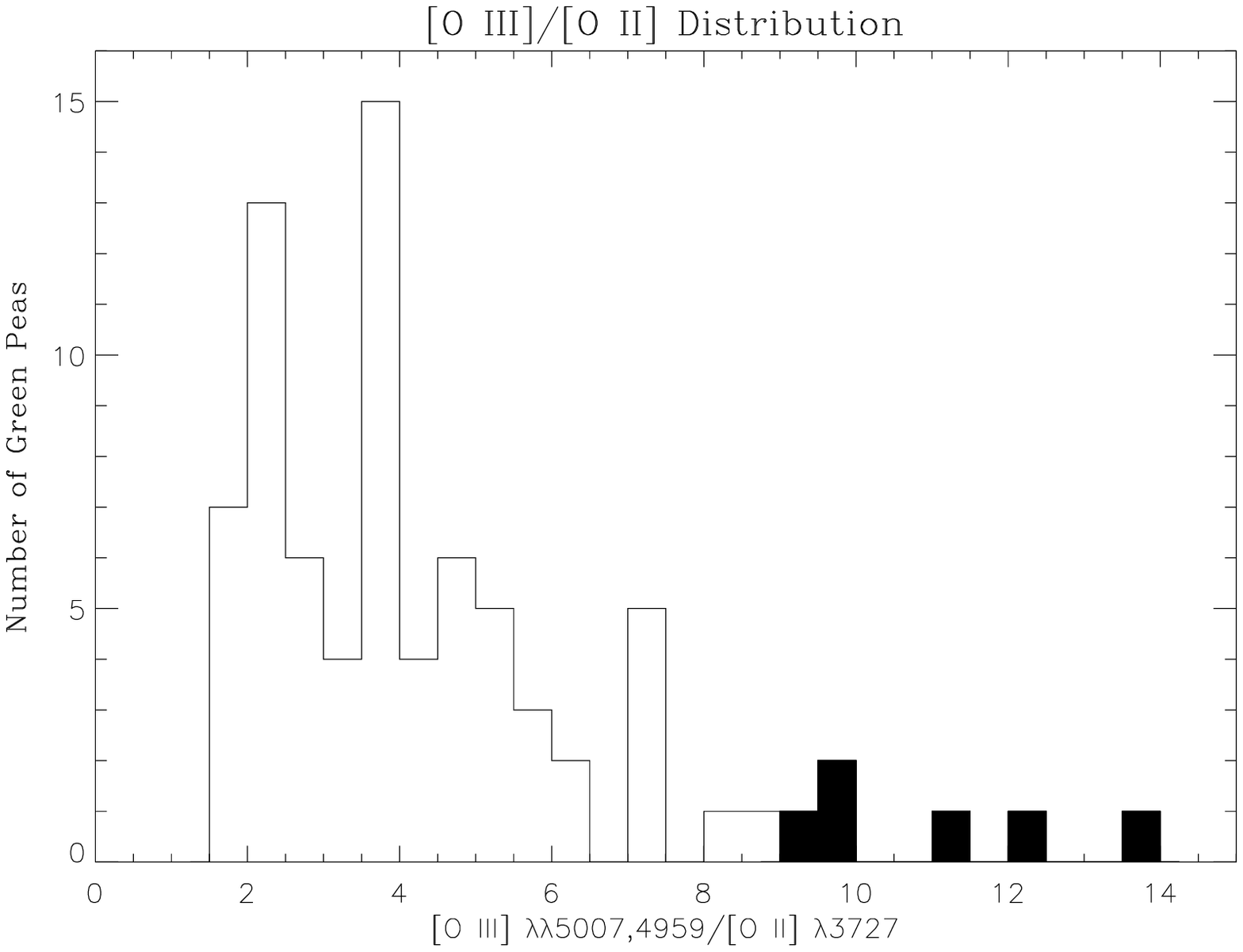}
\caption{A histogram of the \oiii~$\lambda\lambda5007,4959$/\oii~$\lambda$3727 ratios in the Green Peas. The six extreme Peas are shown in black.}
\label{fig_hist}
\end{figure*}

We select the six Peas with the highest \oiii/\oii~ratios for more detailed analysis (Figure~\ref{fig_hist}). These extreme Peas have the greatest potential for a low optical depth or unusual ionizing source. Furthermore, five of these extreme Peas have detectable \heii~$\lambda$4686 emission, which improves our constraints on their ionizing sources (\S\ref{sec:analysis}). Line strengths, errors, and EWs for the six extreme Peas, the Peas with the highest \oiii/\oii~ratios, are listed in Table~\ref{table_lines}. 

\begin{table*}
\vspace*{-0.2in}
\begin{center}
\caption{Selected Emission-Line Intensities of the Extreme Green Peas}
\end{center}
\label{table_lines}
{\scriptsize
\begin{center}
\begin{tabular}{lcccccc}
\hline
& \multicolumn{2}{c}{\peaA} & \multicolumn{2}{c}{\peaB} & \multicolumn{2}{c}{\peaC} \\
\cline{2-3}\cline{4-5}\cline{6-7}
Line & $F(\lambda)/F({\rm H}\beta)$ & $I(\lambda)/I({\rm H}\beta)$ & $F(\lambda)/F({\rm H}\beta)$ & $I(\lambda)/I({\rm H}\beta)$ & $F(\lambda)/F({\rm H}\beta)$ & $I(\lambda)/I({\rm H}\beta)$ \\
\hline
\oii~3727 & 0.655 $\pm$ 0.010 & 0.698 $\pm$ 0.067 & 0.622 $\pm$ 0.006 & 0.703 $\pm$ 0.063 & 0.810 $\pm$ 0.008 & 0.872 $\pm$ 0.077 \\
\hei~3819 & ... & ... & ... & ... & 0.010 $\pm$ 0.002 & 0.010 $\pm$ 0.003 \\
\neiii~3869 & 0.504 $\pm$ 0.009 & 0.532 $\pm$ 0.050 & 0.468 $\pm$ 0.006 & 0.520 $\pm$ 0.046 & 0.535 $\pm$ 0.006 & 0.569 $\pm$ 0.050 \\
H$\delta$~4105 & 0.235 $\pm$ 0.007 & 0.244 $\pm$ 0.023 & 0.246 $\pm$ 0.004 & 0.266 $\pm$ 0.023 & 0.258 $\pm$ 0.004 & 0.270 $\pm$ 0.023 \\
H$\gamma$~4340 & 0.458 $\pm$ 0.008 & 0.468 $\pm$ 0.041 & 0.456 $\pm$ 0.005 & 0.479 $\pm$ 0.040 & 0.463 $\pm$ 0.005 & 0.477 $\pm$ 0.039 \\
\oiii~4363 & 0.134 $\pm$ 0.006 & 0.136 $\pm$ 0.013 & 0.127 $\pm$ 0.003 & 0.133 $\pm$ 0.011 & 0.128 $\pm$ 0.003 & 0.131 $\pm$ 0.011 \\
\niii+\ciii~4640-4650 & ... & ... &  ... & ... & 0.015 $\pm$ 0.004 & 0.015 $\pm$ 0.004 \\
\feiii+\civ~4658 & ... & ... &  ... & ... & 0.009 $\pm$ 0.003 & 0.009 $\pm$ 0.003 \\
\heii~4686 & 0.018 $\pm$ 0.005 & 0.018 $\pm$ 0.006 & 0.020 $\pm$ 0.003 & 0.020 $\pm$ 0.004 & 0.011 $\pm$ 0.003 & 0.011 $\pm$ 0.003 \\
\ariv+\hei~4713 & 0.030 $\pm$ 0.006 & 0.030 $\pm$ 0.006 & 0.019 $\pm$ 0.003 & 0.019 $\pm$ 0.003 & 0.015 $\pm$ 0.003 & 0.015 $\pm$ 0.003 \\
\ariv~4740 & ... & ... & 0.010 $\pm$ 0.003 & 0.010 $\pm$ 0.003 & 0.006 $\pm$ 0.002 & 0.007 $\pm$ 0.002 \\
\oiii~4959 & 2.335 $\pm$ 0.018 & 2.300 $\pm$ 0.188 & 2.216 $\pm$ 0.011 & 2.197 $\pm$ 0.169 & 2.484 $\pm$ 0.012 & 2.471 $\pm$ 0.189 \\
\oiii~5007 & 7.365 $\pm$ 0.053 & 7.235 $\pm$ 0.588 & 6.620 $\pm$ 0.028 & 6.535 $\pm$ 0.499 & 7.479 $\pm$ 0.033 & 7.420 $\pm$ 0.563 \\
\none~5200 & ... & ... & ... & ... & 0.006 $\pm$ 0.002 & 0.006 $\pm$ 0.002 \\
\hei~5876 & 0.114 $\pm$ 0.004 & 0.107 $\pm$ 0.009 & 0.127 $\pm$ 0.003 & 0.117 $\pm$ 0.009 & 0.118 $\pm$ 0.003 & 0.113 $\pm$ 0.008 \\
\oi~6300 & 0.014 $\pm$ 0.004 & 0.013 $\pm$ 0.004 & 0.027 $\pm$ 0.004 & 0.025 $\pm$ 0.004 & 0.025 $\pm$ 0.002 & 0.023 $\pm$ 0.003 \\
\nii~6548 & ... & ... & ... & ... & 0.015 $\pm$ 0.003 & 0.014 $\pm$ 0.003 \\
H$\alpha$~6563 & 3.031 $\pm$ 0.022 & 2.792 $\pm$ 0.201 & 3.162 $\pm$ 0.015 & 2.795 $\pm$ 0.189 & 3.012 $\pm$ 0.014 & 2.796 $\pm$ 0.188 \\
\nii~6584 & 0.049 $\pm$ 0.005 & 0.045 $\pm$ 0.006 & 0.077 $\pm$ 0.006 & 0.068 $\pm$ 0.007 & 0.068 $\pm$ 0.003 & 0.064 $\pm$ 0.005 \\
\sii~6716 & 0.083 $\pm$ 0.006 & 0.075 $\pm$ 0.008 & 0.059 $\pm$ 0.004 & 0.052 $\pm$ 0.005 & 0.070 $\pm$ 0.004 & 0.065 $\pm$ 0.006 \\
\sii~6731 & 0.057 $\pm$ 0.006 & 0.052 $\pm$ 0.007 & 0.066 $\pm$ 0.004 & 0.058 $\pm$ 0.005 & 0.062 $\pm$ 0.004 & 0.057 $\pm$ 0.005 \\
\ariii~7135 & 0.081 $\pm$ 0.008 & 0.073 $\pm$ 0.009 & 0.050 $\pm$ 0.007 & 0.043 $\pm$ 0.007 & 0.061 $\pm$ 0.006 & 0.056 $\pm$ 0.006 \\
$c({\rm H}\beta)$ dex & \multicolumn{2}{c}{0.12} & \multicolumn{2}{c}{0.19} & \multicolumn{2}{c}{0.12} \\
$F({\rm H}\beta)^a$ & 4.61 $\pm$ 0.03 & 6.10 $\pm$ 0.35 & 8.11 $\pm$ 0.03 & 12.67 $\pm$ 0.69 & 8.73 $\pm$ 0.04 & 11.42 $\pm$ 0.62\\
EW(abs)$^b$\AA & \multicolumn{2}{c}{2.23} & \multicolumn{2}{c}{0.00} & \multicolumn{2}{c}{0.00} \\
EW$({\rm H}\beta)$\AA & \multicolumn{2}{c}{224.0} & \multicolumn{2}{c}{246.7} & \multicolumn{2}{c}{334.6} \\
EW$({\rm H}\alpha)$\AA & \multicolumn{2}{c}{964.0} & \multicolumn{2}{c}{1559.0} & \multicolumn{2}{c}{1772.0} \\
\hline
& \multicolumn{2}{c}{\peaD} & \multicolumn{2}{c}{\peaE} & \multicolumn{2}{c}{\peaF} \\
\cline{2-3}\cline{4-5}\cline{6-7}
Line & $F(\lambda)/F({\rm H}\beta)$ & $I(\lambda)/I({\rm H}\beta)$ & $F(\lambda)/F({\rm H}\beta)$ & $I(\lambda)/I({\rm H}\beta)$ & $F(\lambda)/F({\rm H}\beta)$ & $I(\lambda)/I({\rm H}\beta)$ \\
\hline
\oii~3727 & 0.869 $\pm$ 0.005 & 0.986 $\pm$ 0.086 & 0.825 $\pm$ 0.019 & 0.930 $\pm$ 0.118 & 0.807 $\pm$ 0.008 & 0.836 $\pm$ 0.077\\
\hei~3819 & 0.016 $\pm$ 0.003 & 0.018 $\pm$ 0.003 & ... & ... & ... & ... \\
\neiii~3869 & 0.511 $\pm$ 0.004 & 0.569 $\pm$ 0.049 & 0.544 $\pm$ 0.017 & 0.603 $\pm$ 0.076 & 0.376 $\pm$ 0.006 & 0.386 $\pm$ 0.035 \\
H$\delta$~4105 & 0.241 $\pm$ 0.003 & 0.260 $\pm$ 0.022 & 0.267 $\pm$ 0.011 & 0.288 $\pm$ 0.036 & 0.218 $\pm$ 0.004 & 0.221 $\pm$ 0.020 \\
H$\gamma$~4340 & 0.438 $\pm$ 0.003 & 0.461 $\pm$ 0.037 & 0.475 $\pm$ 0.014 & 0.500 $\pm$ 0.059 & 0.426 $\pm$ 0.005 & 0.423 $\pm$ 0.018 \\
\oiii~4363 & 0.115 $\pm$ 0.003 & 0.120 $\pm$ 0.010 & 0.124 $\pm$ 0.010 & 0.130 $\pm$ 0.019 & 0.115 $\pm$ 0.004 & 0.115 $\pm$ 0.010 \\
\feiii$+$\civ~4658 & 0.010 $\pm$ 0.003 & 0.011 $\pm$ 0.003 & ... & ... & 0.032$^c$ $\pm$ 0.005 & 0.031$^c$ $\pm$ 0.006 \\
\heii~4686 & 0.007 $\pm$ 0.002 & 0.008 $\pm$ 0.002 & ... & ... & 0.013 $\pm$ 0.003 & 0.012 $\pm$ 0.003 \\
\ariv+\hei~4713 & 0.021 $\pm$ 0.003 & 0.022 $\pm$ 0.003 & ... & ... & 0.017 $\pm$ 0.003 & 0.017 $\pm$ 0.003 \\
\ariv~4740 & 0.013 $\pm$ 0.003 & 0.013 $\pm$ 0.003 & ... & ... & ... & ... \\
\oiii~4959 & 2.449 $\pm$ 0.010 & 2.417 $\pm$ 0.182 & 2.236 $\pm$ 0.032 & 2.216 $\pm$ 0.241 & 2.001 $\pm$ 0.013 & 1.927 $\pm$ 0.152 \\
\oiii~5007 & 7.392 $\pm$ 0.028 & 7.264 $\pm$ 0.545 & 6.765 $\pm$ 0.090 & 6.679 $\pm$ 0.722 & 6.168 $\pm$ 0.035 & 5.926 $\pm$ 0.464 \\
\none~5200 & ... & ... & ... & ... & ... & ... \\
\hei~5876 & 0.127 $\pm$ 0.002 & 0.116 $\pm$ 0.008 & 0.131 $\pm$ 0.010 & 0.121 $\pm$ 0.015 & 0.114 $\pm$ 0.004 & 0.105 $\pm$ 0.008 \\
\oi~6300 & 0.034 $\pm$ 0.002 & 0.030 $\pm$ 0.003 & ... & ... & 0.016 $\pm$ 0.004 & 0.015 $\pm$ 0.004 \\
\nii~6548 & 0.018 $\pm$ 0.005 & 0.016 $\pm$ 0.004 & ... & ... & ... & ... \\
H$\alpha$~6563 & 3.189 $\pm$ 0.013 & 2.787 $\pm$ 0.185 & 3.193 $\pm$ 0.050 & 2.828 $\pm$ 0.272 & 3.064 $\pm$ 0.020 & 2.786 $\pm$ 0.192 \\
\nii~6584 & 0.069 $\pm$ 0.004 & 0.060 $\pm$ 0.005 & 0.112 $\pm$ 0.024 & 0.099 $\pm$ 0.023 & 0.078 $\pm$ 0.009 & 0.071 $\pm$ 0.009 \\
\sii~6716 & 0.097 $\pm$ 0.004 & 0.084 $\pm$ 0.007 & 0.132 $\pm$ 0.018 & 0.116 $\pm$ 0.019 & 0.063 $\pm$ 0.010 & 0.056 $\pm$ 0.010 \\
\sii~6731 & 0.077 $\pm$ 0.004 & 0.067 $\pm$ 0.005 & 0.079 $\pm$ 0.016 & 0.069 $\pm$ 0.015 & 0.052 $\pm$ 0.010 & 0.047 $\pm$ 0.009 \\
\ariii~7135 & 0.058 $\pm$ 0.004 & 0.049 $\pm$ 0.005 & 0.116 $\pm$ 0.031 & 0.099 $\pm$ 0.028 & 0.044 $\pm$ 0.016 & 0.039 $\pm$ 0.014 \\
$c({\rm H}\beta)$ & \multicolumn{2}{c}{0.21} & \multicolumn{2}{c}{0.19} & \multicolumn{2}{c}{0.11} \\
$F({\rm H}\beta)^a$ & 9.06 $\pm$ 0.03 & 14.63 $\pm$ 0.79 & 1.74 $\pm$ 0.02 & 2.70 $\pm$ 0.21 & 6.71 $\pm$ 0.04 & 8.87 $\pm$ 0.49 \\
EW(abs)$^b$\AA & \multicolumn{2}{c}{0.84} & \multicolumn{2}{c}{0.00} & \multicolumn{2}{c}{4.67} \\
EW$({\rm H}\beta)$\AA & \multicolumn{2}{c}{228.6} & \multicolumn{2}{c}{184.9} & \multicolumn{2}{c}{140.5} \\
EW$({\rm H}\alpha)$\AA & \multicolumn{2}{c}{1158.0} & \multicolumn{2}{c}{1086.0} & \multicolumn{2}{c}{812.5} \\
\hline
\end{tabular}
\end{center}
\raggedright{
The first 22 rows list the fluxes relative to H$\beta$~for the designated emission lines. For each object, we list the observed fluxes ($F(\lambda)$) and the fluxes corrected for reddening and stellar absorption ($I(\lambda)$).\\
$^a$We list both the observed flux and the flux corrected for reddening and stellar absorption. Fluxes are in $10^{-15}$ erg s$^{-1}$ cm$^{-2}$.\\
$^b$Best-fit stellar absorption equivalent width. \\
$^c$Includes nebular \feiii~$\lambda$4658 emission and broad \civ~emission. \\
}
\break
}
\end{table*}

The Peas have high electron temperatures, as expected for low-metallicity objects. After correcting for reddening and stellar absorption as discussed above, we re-calculated the electron temperature. We then calculated the electron density from the observed \sii~$\lambda$6716/$\lambda6731$ ratio with {\tt temden} and iterated until both temperature and density converged. The revised temperatures are still consistent with the temperatures assumed in the reddening calculations. Our calculated temperatures and densities and the metallicity measurements of \citet{izotov11} are listed in Table~\ref{table_neb} for the extreme Peas. \citet{izotov11} determined abundances using the direct method, which calculates the oxygen abundance given the observed \oiii~and \oii~fluxes and the electron temperature derived from the \oiii~($\lambda4959+\lambda5007$)/$\lambda$4363 ratio. Table~\ref{table_neb} also lists the SFR, derived from the \halpha~luminosity according to
\begin{equation}
\label{eqn_sfr}
\log SFR =\log L(H\alpha)-41.27,
\end{equation}where SFR is in units of \Msol yr$^{-1}$ and $L(H\alpha)$ is in \ergps \citep{kennicutt12}. Our SFRs are higher than the ones found for these objects by \citet{cardamone09}, despite our use of the \citet{kennicutt12} relation, which leads to lower SFRs relative to the \citet{kennicutt98} relation. We find \halpha~luminosities for the extreme Peas that are typically a factor of 2 greater than found by \citet{cardamone09} due to our different treatment of reddening. \citet{cardamone09} assume an intrinsic H$\alpha$/H$\beta$~ratio appropriate for $T_e=10^4$ K for all the Peas, whereas the six extreme Peas have temperatures of $\sim$15000 K (Table~\ref{table_neb}). In addition, \citet{cardamone09} use the \citet{cardelli89} reddening law, instead of the \citet{calzetti00} law we assume.

\begin{table*}
\vspace*{-0.2in}
\begin{center}
\caption{Properties of the Extreme Green Peas}
\label{table_neb}
{\scriptsize
\begin{tabular}{lccccccc}
\hline
SDSS ID & z$^a$ & \oiii~$\lambda\lambda4959,5007/\lambda4363$ & \sii~$\lambda6716/\lambda6731$ & $T_e$ & $n_e$ & $Z^b$ & SFR$^c$ \\ 
& & & & (K) & (cm$^{-3}$) &  (\Zsol) & (\Msol yr$^{-1}$) \\
\hline
\peaA & 0.1410 & 69.901 $\pm$ 6.295 & 1.439$^d$ $\pm$ 0.204 & 15069$^e$ & $\leq$206$^f$ & 0.204 & 4.9 $\pm$ 0.2\\
\peaB & 0.1957 & 65.473 $\pm$ 5.145 & 0.897 $\pm$ 0.096 & 15507 & 1088 & 0.158 & 21.0 $\pm$ 0.8\\
\peaC & 0.1439 & 75.320 $\pm$ 5.886 & 1.132 $\pm$ 0.110 & 14566 & 378 & 0.209 & 9.6 $\pm$ 0.4\\
\peaD & 0.1488 & 80.505 $\pm$ 6.121 & 1.259 $\pm$ 0.108 & 14156 & 171 & 0.224 & 13.2 $\pm$ 0.5\\
\peaE & 0.2219 & 68.501 $\pm$ 9.140 & 1.680$^d$ $\pm$ 0.433 & 15209$^e$ & $\leq$189$^f$ & 0.166 & 6.0 $\pm$ 0.3\\
\peaF & 0.1650 & 68.502 $\pm$ 5.778 & 1.193 $\pm$ 0.301 & 15204 & 272 & 0.148 & 10.1 $\pm$ 0.4\\
\hline
\end{tabular}
\flushleft{
$^a$From SDSS DR7. \\
$^b$From \citet{izotov11}. We assume a solar metallicity of 12+log(O/H)=8.69 \citep{allende01}.\\
$^c$Using the conversion from H$\alpha$~luminosity given in \citet{kennicutt12}. \\
$^d$The \sii~ratio is at the low-density limit, and we are unable to determine a density.\\
$^e$Calculated assuming a density of 100 cm$^{-3}$.\\
$^f$The upper limit is calculated from the 1$\sigma$~errors for the measured \sii~ratio.\\
}
\break
}
\end{center}
\end{table*}

As evident in Table~\ref{table_neb}, the extreme Peas have an average metallicity near 0.2\Zsol. At metallicities $Z \leq 0.2$\Zsol,  temperature gradients in nebulae can cause the direct method to under- or over-estimate metallicities by 0.2-0.5 dex \citep{lopez12}. \oii~temperature estimates are necessary to correct for this effect, but unfortunately, the \oii~$\lambda\lambda$7319 and 7330 lines are dominated by noise in the SDSS spectra. This metallicity uncertainty should be kept in mind for the extreme Peas \peaBs, \peaEs, and \peaFs, which have metallicities $< 0.2$\Zsol. 

To identify weak lines in the Peas' spectra, we created stacked spectra for several subsets of the Peas. We first corrected the spectra for redshift using the IRAF task {\tt dopcor}. We then normalized the spectra to the mean flux in the 4200-4300 \AA~range, which is free of strong spectral lines. As a check, we also created stacked spectra using the SDSS continuum-subtracted spectra and no normalization. These additional spectra had a flat continuum, making the hydrogen absorption lines more easily visible. Aside from this difference, using the continuum-subtracted spectra does not change our results. Following the normalization, we re-binned the spectra to a dispersion of 1 \AA/pixel and averaged the spectra, rejecting pixels that deviated by more than 5$\sigma$ from the median. We created stacked spectra for the full sample of 78 star-forming Peas with \oiii~$\lambda5007 > $\oii~$\lambda$3727 and for a subset of 53 Peas with the strongest \oiii~emission. Although our analysis focuses on the six extreme Peas, we created an additional stacked spectrum of the 15 highest \oiii/\oii~ratio Peas to improve the S/N. 

\subsection{Equivalent Widths and Age Constraints}
\label{sec:age}

Although several studies have proposed an `aged starburst' hypothesis to explain the high ionization conditions in the Peas and similar galaxies \citep[e.g.,][]{overzier09}, a number of spectral features suggest younger ages, particularly for the highest ionization Peas. Tables~\ref{table_age} and~\ref{table_age_cont} compare the EWs and corresponding age estimates for the extreme Peas. The Balmer line EWs are large; the full sample of 78 Peas has an average redshift-corrected H$\alpha$ EW of 607 \AA, while the high-ionization Peas in Table~\ref{table_age}~have H$\alpha$ EWs $\sim$100-1000 \AA~greater than this average. To estimate the Peas' ages, we ran Starburst99 models \citep{leitherer99} at $Z=0.2$\Zsol~for a $10^6$\Msol~instantaneous burst and for continuous star formation with the SFRs in Table~\ref{table_neb}. We used a \citet{kroupa01} initial mass function from 0.1-150\Msol, the Padova evolutionary tracks and mass-loss rates with AGB stars \citep{vazquez05}, and the \citet{pauldrach01} and \citet{hillier98} stellar atmospheres. For the Peas' H$\alpha$~EWs, the instantaneous burst model places upper limits on the burst ages of 4.2-5.1 Myr (Table~\ref{table_age}). The H$\beta$ EWs suggest slightly younger ages, with upper limits of 3.7-4.9 Myr. These latter values may be more reliable due to the low continuum levels in the H$\alpha$~spectral region and blending with the \nii~lines. We note, however, that both the H$\beta$ and H$\alpha$ EWs give similar ages.

\begin{table*}
\vspace*{-0.2in}
\begin{center}
\caption{Ages of the Extreme Green Peas (Instantaneous Burst)}
\label{table_age}
{\scriptsize
\begin{tabular}{lcccccccccc}
\hline
Galaxy & \multicolumn{2}{c}{H$\alpha$} & & \multicolumn{2}{c}{H$\beta$} & & \multicolumn{2}{c}{\hei~$\lambda$3819} & & \\
\cline{2-3} \cline{5-6} \cline{8-9}
& EW$^a$ & Age$^b$ & & EW$^a$ & Age$^b$ & & EW$^a$ & Age$^c$ & M$_{\rm Burst}$(3 Myr)$^d$ & M$_{\rm Burst}$(H$\beta$ EW)$^e$\\ 
& (\AA) & (Myr) & & (\AA) & (Myr) & & (\AA) & (Myr) & ($10^6$\Msol) & ($10^6$\Msol) \\
\hline
\peaAs & 845 & $\leq$ 4.9 &  &196 & $\leq$ 4.4 & &... & ... & 13.5 & 39.9\\
\peaBs & 1304 & $\leq$ 4.5 &  &206 & $\leq$ 4.4 & & ... & ... & 57.9 & 170.7\\
\peaCs & 1550 & $\leq$ 4.2 &  &293 & $\leq$ 3.7 & &1.91 & $< 2$ & 26.5 & 42.0 \\
\peaDs & 1008 & $\leq$ 4.8 &  &199 & $\leq$ 4.4 & &1.99 & $< 2$ & 36.4 & 107.4 \\
\peaEs & 889 & $\leq$ 4.9 &  &151 & $\leq$ 4.7 & &... & ... & 16.6 & 64.4 \\
\peaFs & 697 & $\leq$ 5.1 &  &121 & $\leq$ 4.9 & &... & ... & 27.7 & 135.6 \\
\hline
\end{tabular}
\flushleft{
$^a$Redshift-corrected. \\
$^b$Age using the instantaneous burst Starburst99 model. \\
$^c$Age using the instantaneous burst model of \citet{gonzalez99}. \\
$^d$From the H$\alpha$~luminosity, assuming an age of 3 Myr. \\
$^e$From the H$\alpha$~luminosity, using the H$\beta$ EW-derived age from Column 4. \\
}
\break
}
\end{center}
\end{table*}

\begin{table*}
\vspace*{-0.2in}
\begin{center}
\caption{Ages of the Extreme Green Peas (Continuous Star Formation)}
\label{table_age_cont}
{\scriptsize
\begin{tabular}{lcccccccc}
\hline
Galaxy & \multicolumn{2}{c}{H$\alpha$} & & \multicolumn{2}{c}{H$\beta$} & & \multicolumn{2}{c}{\hei~$\lambda$3819}\\
\cline{2-3} \cline{5-6} \cline{8-9}
& EW$^a$ & Age$^b$ & & EW$^a$ & Age$^b$ & & EW$^a$ & Age$^c$ \\ 
& (\AA) & (Myr) & & (\AA) & (Myr) & & (\AA) & (Myr) \\
\hline
\peaAs & 845 & $\leq$ 37.4 &  &196 & $\leq$ 16.0 & & ... & ...\\
\peaBs & 1304 & $\leq$ 13.5 &  &206 & $\leq$ 13.9 & & ... & ...\\
\peaCs & 1550 & $\leq$ 9.7 &  &293 & $\leq$ 7.5 &  &1.91 & $< 3 $ \\
\peaDs & 1008 & $\leq$ 25.5 &  &199 & $\leq$ 15.3 &  &1.99 & $< 3 $\\
\peaEs & 889 & $\leq$ 33.4 &  &151 & $\leq$ 31.9 &  &... & ... \\
\peaFs & 697 & $\leq$ 59.3 &  &121 & $\leq$ 58.6 &  &... & ... \\
\hline
\end{tabular}
\flushleft{
$^a$Redshift-corrected. \\
$^b$Age assuming the continuous star formation Starburst99 model. \\
$^c$Age from the \citet{gonzalez99} continuous star formation model. \\
}
\break
}
\end{center}
\end{table*}

\begin{figure*}
\epsscale{1.0}
\plotone{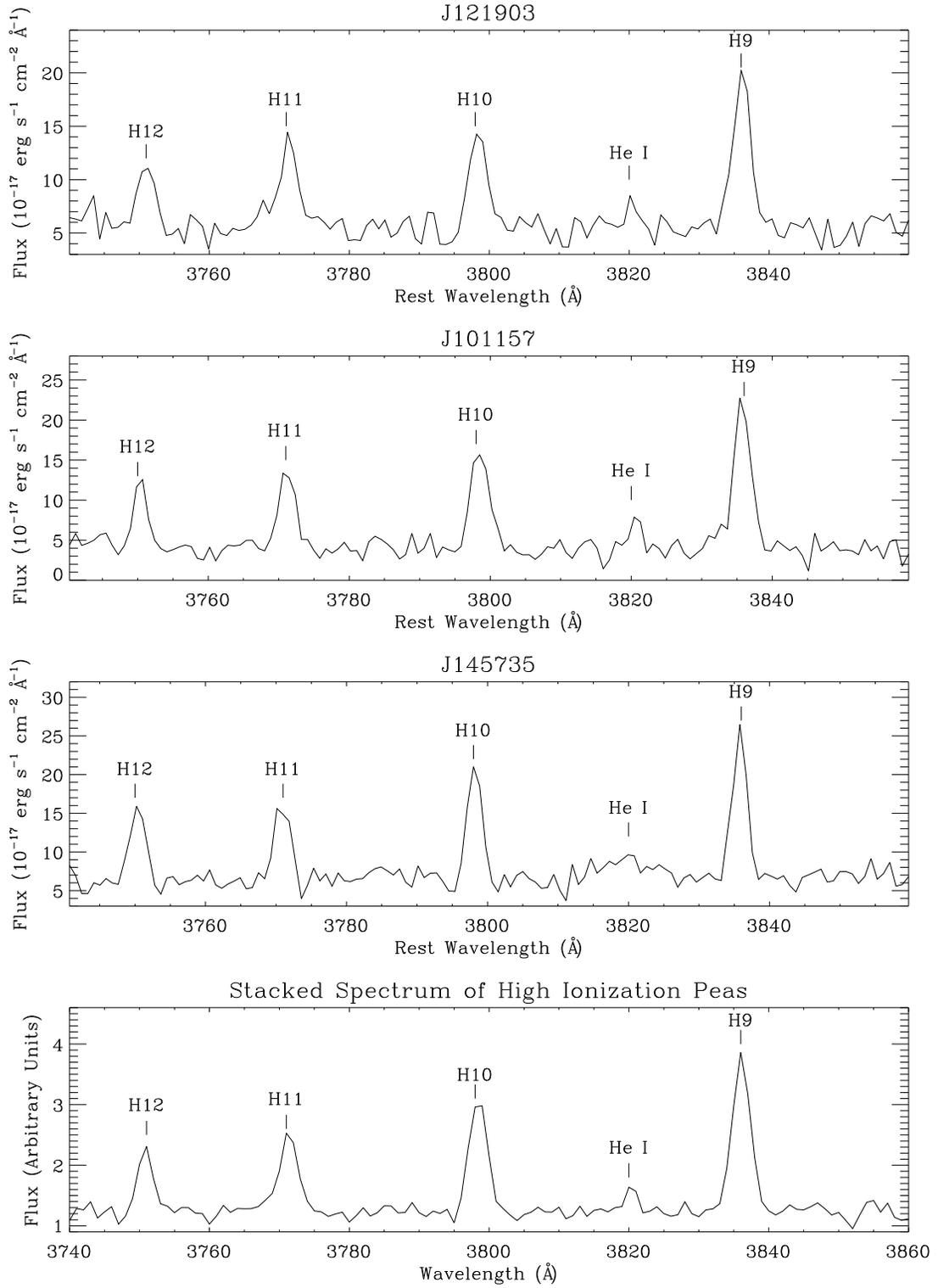}
\caption{The top three panels show the 3740-3860 \AA~region for Peas \peaBs, \peaCs, and \peaDs. \peaCs and \peaDs have detections of the weak \hei~$\lambda$3819 line in emission, and \hei~$\lambda$3819 may be present in \peaBs as well. The bottom panel shows the same region for the stacked spectrum of the 15 Peas with the highest \oiii/\oii~ratios.}
\label{fig_hei}
\end{figure*}

An additional indication of the Peas' young ages comes from the detection of the \hei~$\lambda$3819 line. \citet{gonzalez99} show that this weak line is easily covered up by stellar absorption and is only present for the first 3 Myr following an instantaneous burst of star formation. The \hei~$\lambda$3819 line is evident in the individual spectra of 5 of the 78 Peas, including Peas \peaCs and \peaDs, which are among the extreme Peas (Figure~\ref{fig_hei}). Pea \peaBs~may also exhibit \hei~$\lambda$3819 emission, but the emission is only present at a 2.6$\sigma$ level. The $\lambda$3819 line shows up clearly in each of the stacked spectra as well (Figure~\ref{fig_hei}). 

The H Balmer lines in the stacked spectra also point to a young age. As with the \hei~$\lambda$3819 line, stellar absorption quickly dominates the higher order Balmer lines \citep{gonzalez99}. The stacked spectra show the entire Balmer series in emission, which again indicates an upper limit of 3 Myr on the age of an instantaneous burst \citep{gonzalez99}.

If the star formation is continuous, however, the Peas' ages may be older than 3 Myr. In the case of continuous star formation Starburst99 models, the Balmer line EWs give upper age limits of 7-60 Myr (Table~\ref{table_age_cont}). Similarly, the Balmer emission in the stacked spectra gives an upper age limit of 20 Myr. The observed \hei~$\lambda$3819 EWs indicate young ages, however, even with continuous star formation. Although \hei~$\lambda$3819 can exist in emission for the first 10 Myr of continuous star formation, the observed \hei~$\lambda$3819 EWs imply ages of 3-6 Myr \citep{gonzalez99}. 

Table~\ref{table_age} also lists the stellar mass formed in the burst. To estimate the size of the burst, we compare the observed H$\alpha$ luminosities with the ionizing photon production rate, $Q({\rm H})$, in the Starburst99 instantaneous burst model. We convert $Q({\rm H})$ to an H$\alpha$ luminosity using the Balmer line ratios, Case B emissivities, and Case B recombination coefficients at 15,000 K from \citet{storey95}. We calculate the starburst mass twice: once assuming an age of 3 Myr, as suggested by the \hei~$\lambda$3819 emission, and once with the instantaneous burst age from the H$\beta$~EW. Since the H$\alpha$~luminosity is roughly constant for the first 3 Myr, the burst mass corresponding to an age of 3 Myr should also be valid for younger ages. As shown in Table~\ref{table_age}, the derived burst masses range from 14-171 million \Msol. Using the Peas' total stellar masses from \citet{izotov11} and the 3 Myr ages, the bursts account for 2-26\% of the total stellar mass of the Peas. If the older ages of 4-5 Myr are adopted, the fraction of the total mass increases. While the burst fractions for four of the extreme Peas remain in the 2-26\% range, Peas \peaBs~and \peaEs~have burst fractions of $76\%$ and $87\%$ if the older ages are used. Such high fractions may not be realistic, however. In particular, given the detection of \hei~$\lambda$3819 emission in \peaBs, the younger age and lower burst fraction are preferable. 

In either case, continuous or instantaneous star formation, the Peas are evidently young and powerful starbursts. We now turn to the question of the different ionizing sources that may be present in these objects and whether they can explain the observed emission. The Peas themselves may span a range of ages, and different sources may dominate the ionization in individual galaxies.


\section{High Ionization Lines and Possible Origins}
\label{sec:analysis}

\begin{figure*}
\epsscale{1.0}
\plottwo{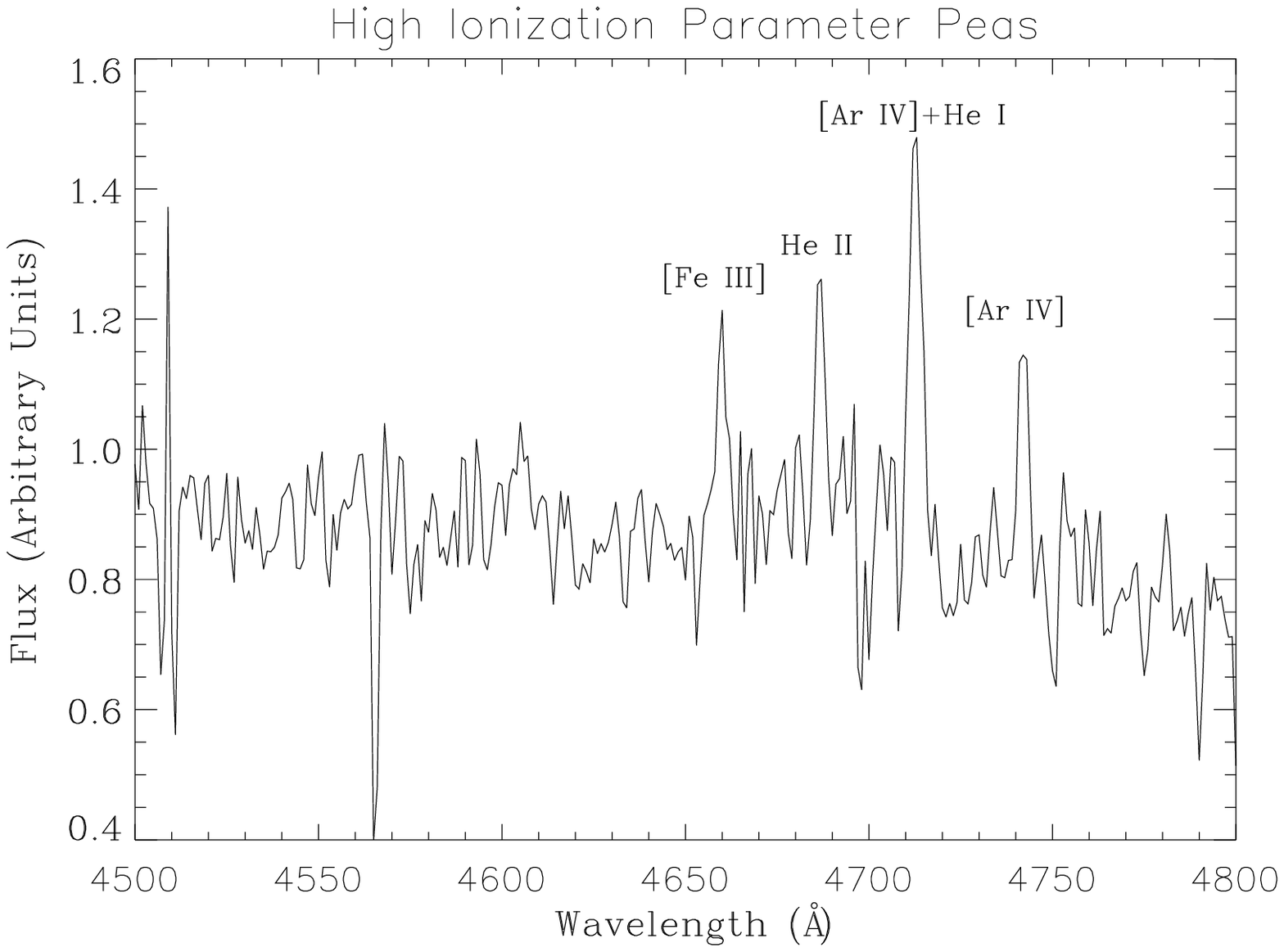}{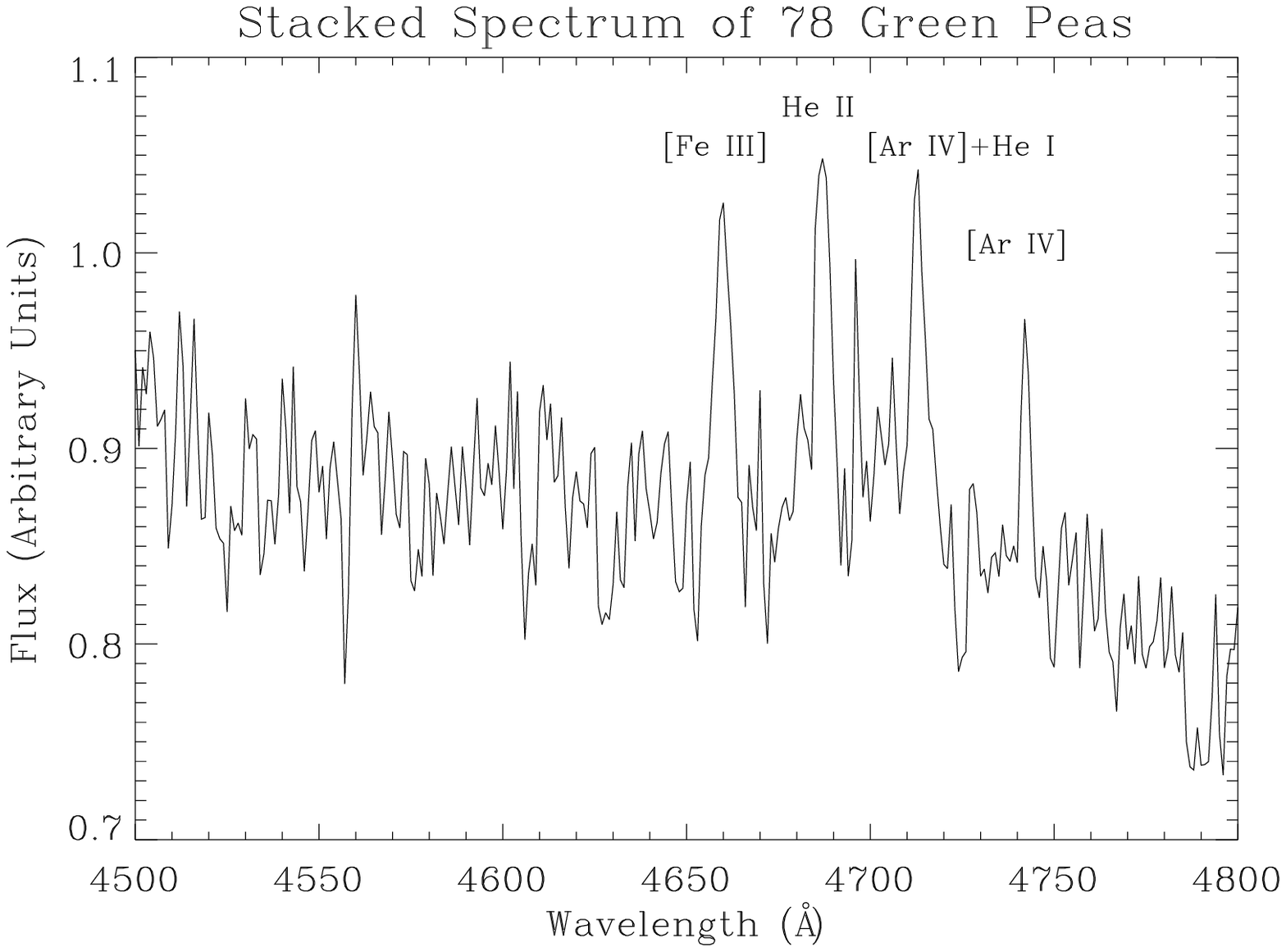}
\caption{The left panel shows the 4500-4800 \AA~region of the stacked spectrum for the 15 Peas with the highest \oiii/\oii~ratios. The right panel displays the same region for a stacked spectrum of all 78 star-forming Peas. The nebular \feiii~$\lambda$4658, \heii~$\lambda$4686, \ariv+\hei~$\lambda$4713, and \ariv~$\lambda$4740 lines are labelled. No broad WR features are visible.}
\label{fig_stack}
\end{figure*}

\begin{figure*}
\epsscale{1.0}
\plotone{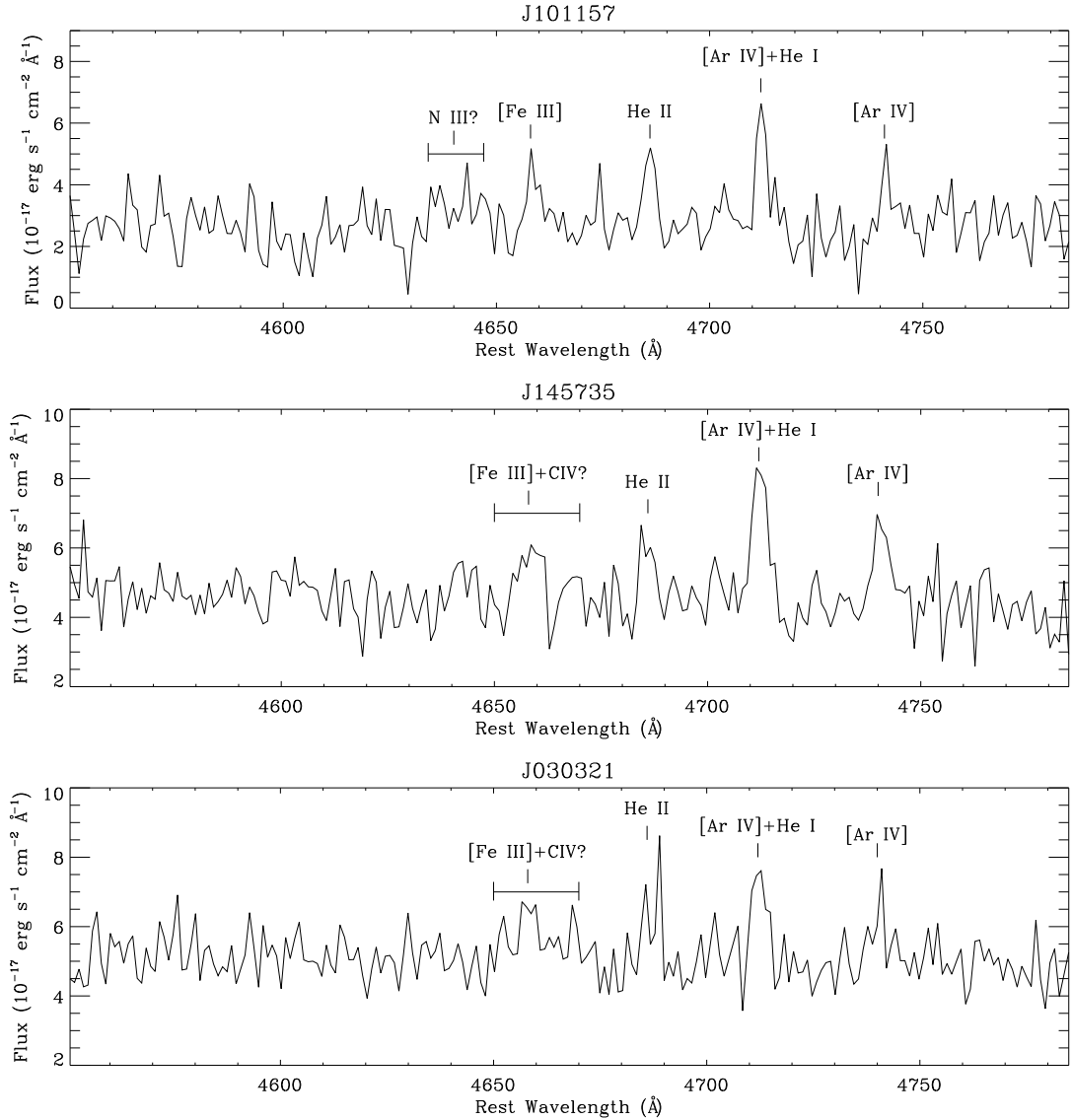}
\caption{The WR blue bump and \heii~emission for the three Peas which may have WR features.}
\label{fig_bb}
\end{figure*}

The Green Peas have \oiii~$\lambda\lambda$5007,4959/\oii~$\lambda$3727 ratios ranging from 1.5 to as high as 13.7, which may indicate an extremely high ionization parameter and/or a low optical depth. Several of the Peas show other high ionization lines in their spectra, with strong \neiii~$\lambda$3869 emission as well as detectable \ariv~$\lambda$4712, \ariv~$\lambda$4740, and \heii~$\lambda$4686. In particular, we detect \heii~emission in 5 of the 6 most extreme (i.e., highest \oiii/\oii~ratio) Peas. The presence of these weak, high ionization lines suggests the existence of a hard ionizing radiation field, such as that produced by WR stars. Interestingly, while the high-ionization lines show up clearly in the stacked spectra of the Peas (Figure~\ref{fig_stack}), no stellar WR features are visible. In particular, the blue bump at 4640-4650~\AA, red bump at 5808~\AA, and broad \heii~$\lambda$4686 component are absent. Instead, the \heii~emission line appears narrow, suggesting a nebular origin. In the Peas' individual spectra, the S/N is lower, but WR features are tentatively present in three of the extreme Peas (Figure~\ref{fig_bb}). 

Some or all of the Peas may therefore fall into a mysterious class of galaxies with \heii~emission but no WR features \citep{guseva00, thuan05, shirazi12}. \citet{shirazi12} have recently analyzed 2865 \heii-emitting galaxies in SDSS. They find that the fraction of \heii-emitters without WR features systematically increases with decreasing metallicity \citep{shirazi12}. Possible origins for the \heii~emission include WR stars with weak stellar lines, chemically homogeneous massive stars, AGNs, high-mass X-ray binaries (HMXBs), and fast, radiative shocks \citep{shirazi12}. The precise cause of the \heii~emission may vary from galaxy to galaxy, however. Here, we specifically evaluate these proposed ionizing sources for the six extreme Peas. We note that if the Peas are indeed optically thin, their low optical depths would further enhance their \heii/H$\beta$~ratios by decreasing the observed H$\beta$ emission. We therefore use both density-bounded and radiation-bounded models to evaluate the possible explanations for the \heii~emission and line ratios observed in the most extreme Green Peas. 

In Table~\ref{table_he2}, we list the \heii~luminosities and the WR blue bump luminosities or 1$\sigma$ limits for the 6 extreme Peas. We calculate the 1$\sigma$~blue bump and \heii~detection limits using Equation~\ref{error_eqn}, correcting for reddening, and assuming rest-frame line widths of 75 \AA~and 7.07 \AA, respectively. The \heii~width of 7.07 \AA~is the average rest-frame width of the detected $\lambda$4686 lines. The adopted blue bump width corresponds to the maximum extent of the blue bump seen in the composite WN spectra of \citet{crowther06}. As the actual extent of the blue bump may be less than this value \citep[see e.g.,][]{fernandes04}, these limits are likely overestimates. In addition, we calculate the detection limits for the WR \siiii~$\lambda$4565 line by assuming it has the same width as the \heii~line. Any proposed explanation for the high ionization emission must be consistent with the ages and starburst masses of the galaxies and these constraints on the WR and \heii~emission.

\begin{table*}
\vspace*{-0.2in}
\begin{center}
\caption{\heii~and WR Emission in the Extreme Green Peas}
\label{table_he2}
{\scriptsize
\begin{tabular}{lccccc}
\hline
Galaxy & $L({\rm H}\alpha)$ & $L($\heii$~\lambda4686)$ & Blue Bump $L($\niii$~\lambda4640)$ & $L($\siiii$~\lambda4565)^a$ & $L($\civ$~\lambda4658)^b$ \\ 
& ($10^{42}$\ergps) & ($10^{39}$\ergps) & ($10^{39}$\ergps) & ($10^{39}$\ergps) & ($10^{39}$\ergps) \\
\hline
\peaAs & 0.91 & 5.87 & $\leq 4.77^a$ & $\leq 1.45$ & ...\\
\peaBs & 3.91 & 28.7 & $\leq 11.2^a$ & $\leq 3.41$ & ...\\
\peaCs & 1.79 & 7.13 & 9.79 & $\leq 1.98$ & ...\\
\peaDs & 2.46 & 6.64 & $\leq 7.03^a$ & $\leq 2.14$ & 9.32 \\
\peaEs & 1.12 & $\leq 4.12^a$ & $\leq 13.4^a$ & $\leq 4.08$ & ... \\
\peaFs & 1.87 & 8.24 & $\leq 5.79^a$ & $\leq 1.76$ & 20.8 \\
\hline
\end{tabular}
\flushleft{
$^a$1$\sigma$ limit. \\
$^b$May include nebular \feiii~emission.\\
\break
}
}
\end{center}
\end{table*}

\subsection{Stellar WR Feature}
\label{sec:stellar}

Although the narrow width of the observed \heii~$\lambda$4686 line seems to indicate a nebular origin, a narrow stellar line could contribute some portion of the emission. In particular, while most WR stars have broad \heii~emission lines, the WN9 subtype has a \heii~$\lambda$4686 full width at half maximum (FWHM) of only $~6\pm2$\AA~\citep{chandar03}. Our observed \heii~FWHMs range from $\sim3-5$ \AA, slightly lower than the WN9 FWHMs. For comparison, the SDSS spectral resolution at these wavelengths is $\sim2-3$ \AA, which suggests these lines are unresolved and not stellar in origin. Given the low S/N of the \heii~line and the uncertainty in the WN9 FWHM, however, we cannot rule out the WN9 stellar line hypothesis. Furthermore, observations suggest that late-type WR stars may have weaker or narrower lines at lower metallicities \citep{conti89,crowther06}.

Assuming the \heii~is purely stellar in origin, we derive the number of WN9 stars necessary to produce the observed \heii~line luminosities. We adopt the $\lambda$4686 luminosities for late-type WN (WNL) stars given in \citet{brinchmann08}: $L(\lambda4686)= 4.3 \times 10^{35}$\ergps for $Z < 0.2$\Zsol~and $L(\lambda4686)=2.47 \times 10^{36}$\ergps~for $Z \geq 0.2$\Zsol. Since the Peas have average metallicities of $\sim0.2$\Zsol, we calculate the number of WNL stars using both luminosities and list the derived numbers in Table~\ref{table_wn9}. Brinchmann et al. derived the $\lambda$4686 luminosities from WN5-WN6 stars, which are more well studied than WN9 stars. However, WN5-6 stars generally have stronger \heii~emission than WN9 stars \citep[e.g.,][]{crowther06}. The mean \heii~$\lambda$4686 luminosity of WN5-WN6 stars in the Large Magellanic Cloud (LMC) is a factor of 2.4 greater than that of WN7-WN9 stars \citep{crowther06}. We therefore note that the required number of WN9 stars may be underestimated. 

\begin{table*}
\vspace*{-0.2in}
\begin{center}
\caption{Constraints on WN9 Stars in the Green Peas}
\label{table_wn9}
{\scriptsize
\begin{tabular}{lccccccc}
\hline
Galaxy & \multicolumn{2}{c}{$N$(WN9 stars)$^a$} & \multicolumn{2}{c}{$N$(WN9 stars)$^b$} & $N$(WN9 stars)$^c$& \multicolumn{2}{c}{WN9/O} \\ 
& \multicolumn{2}{c}{from \heii~$\lambda$4686} & \multicolumn{2}{c}{from \niii~$\lambda$4640} & from \siiii~$\lambda$4565 & \multicolumn{2}{c}{from \heii~$\lambda$4686} \\
& $Z < 0.2$\Zsol & $Z \geq 0.2$\Zsol & $Z < 0.2$\Zsol & $Z \geq 0.2$\Zsol & & $Z < 0.2$\Zsol & $Z \geq 0.2$\Zsol \\
\hline
\peaAs & 13600 & 2380 & $\leq 11900^d$ & $\leq 7570^d$ & $\leq 2200$ & 0.12-0.30 & 0.02-0.05 \\
\peaBs & 66700 & 11600 & $\leq 28000^d$ & $\leq 17800^d$ & $\leq 5170$ & 0.14-0.34 & 0.02-0.06 \\
\peaCs & 16600 & 2890 & 24500 & 15500 & $\leq 3000$ & 0.13-0.19 & 0.02-0.03 \\
\peaDs & 15400 & 2690 & $\leq 17600^d$ & $\leq 11200^d$ & $\leq 3240$ & 0.05-0.13 & 0.009-0.02 \\
\peaFs & 19200 & 3340 & $\leq 14500^d$ & $\leq 9190^d$ & $\leq 2670$ & 0.06-0.21 & 0.01-0.04 \\
\hline
\end{tabular}
\flushleft{
$^a$Number of stars needed to produce the observed \heii~luminosity.\\
$^b$Limit on the WN9 population from \niii~$\lambda$4640.\\
$^c$1$\sigma$ limits on the WN9 population from \siiii~$\lambda$4565.\\
$^d$From the 1$\sigma$ limit. \\
}
\break
}
\end{center}
\end{table*}

In addition to the blue bump emission at $\lambda$4640 from \niii, WN9 stars show \siiii~$\lambda$4565 emission \citep{guseva00}. Based on the blue bump and \siiii~luminosities or 1$\sigma$ detection limits (Table~\ref{table_he2}), we calculate the number of WN9 stars that could be present and list these values in Table~\ref{table_wn9}. We adopt the WNL blue bump luminosities of \citet{brinchmann08}, which are $4.0 \times 10^{35}$ \ergps for $Z < 0.2$\Zsol~and $6.3 \times 10^{35}$\ergps~for $Z \geq 0.2$\Zsol. \citet{schaerer98} list WNL star blue bump luminosities of $\sim2 \times 0^{36}$\ergps. Adopting these higher values would result in significantly lower estimates of the number of WNL stars present. For the \siiii~$\lambda$4565 luminosity, we adopt $L(\lambda4565)= 6.6 \times 10^{35}$\ergps~from \citet{guseva00}. We note that each of these line luminosities may be uncertain by a factor of two \citep{crowther06}. Even within a given WN subtype, \heii~and blue bump luminosities span a wide range of values \citep{crowther06}, and \citet{guseva00} admit that the \siiii~$\lambda$4565 luminosities are poorly known. 

Assuming the $Z < 0.2$\Zsol~values, the extreme Peas would need to contain 13,000-68,000 WN9 stars to account for the observed $\lambda$4686 luminosities (Table~\ref{table_wn9}). As in \S\ref{sec:age}, we compute the number of O stars in the Peas by comparing the observed H$\alpha$~luminosities to the ionizing photon rates in Staburst99 models. We determine how many 10$^6$ \Msol~bursts would generate the observed H$\alpha$ luminosities and scale the number of O stars in the Starburst99 models accordingly. We note that this approach accounts for the WR contribution to the H$\alpha$ emission. As before, we use both the H$\beta$ EW-derived ages and an age of 3 Myr. From the estimated numbers of O stars, the required low-metallicity WN9 population indicates WR/O ratios of 0.05-0.34 (Table~\ref{table_wn9}). In all but one case, these high WR/O ratios are impossible to match with either the instantaneous burst or continuous star formation Starburst99 models. Furthermore, while the blue bump limits are consistent with this number of WN9 stars at the 3$\sigma$ level, the \siiii~limits do not allow such a large population of WN9 stars. 

A population of WN9 stars with the $Z \geq 0.2$\Zsol~luminosities could be possible. The Peas would need 2,000-12,000 WN9 stars, resulting in WR/O ratios of 0.01-0.06. The Starburst99 models show that these WR/O ratios occur in an instantaneous burst between 3 and 5 Myr in age. The 1$\sigma$ blue bump and \siiii~$\lambda$4565 detection limits are also compatible with this number of WN9 stars. As WNL stars may exist on the main sequence \citep[e.g.,][]{martins08, martins09, grafener11}, younger ages are also not in conflict with the existence of a WNL population.

Although the presence of sufficient WN9 stars is possible, it is by no means certain. The feasibility of the WN9 scenario depends strongly on the assumed emission line luminosities. To account for the observed \heii~emission, the WN9 stars in the Peas must have the high \heii~luminosities characteristic of stars with $Z \geq 0.2$\Zsol. While the exact metallicity dependence of WNL line luminosities is not yet understood, the lower line luminosities associated with metal-poor stars are probably more appropriate for the Peas. Even these latter luminosities may be higher than is realistic, however. As discussed above, the adopted WN9 \heii~line luminosities may be overestimated by a factor of 2-3. Finally, none of the higher S/N stacked spectra show any evidence of WR features (Figure~\ref{fig_stack}), although the blue bump emission in the low-metallicity case should be almost as strong as \heii~$\lambda$4686. Given the significant uncertainties in the WN9 luminosities and line widths \citep[e.g.,][]{crowther06}, we cannot definitively rule out the WN9 scenario. However, we conclude that stellar emission from WN9 stars is unlikely to be the dominant source of the observed \heii~emission.

\subsection{Ionization by Hot Stars}
\label{sec:nebular}

\subsubsection{WR Stars}
\label{sec:nebular:wr}

The other possibility is that the \heii~is nebular in origin rather than stellar. The clear presence of other high-ionization nebular lines, such as \ariv~$\lambda\lambda$4712,4740 and strong \neiii~$\lambda$3869, supports this scenario.  If the nebular \heii~emission is due to photoionization from stars, the stars responsible must have sufficient emission shortward of 228 \AA. Extremely hot WR stars \citep[e.g.,][]{smith02} or O stars \citep{kudritzki02} may have spectral energy distributions hard enough to photoionize \heii. 

To test this hypothesis, we create a grid of photoionization models with CLOUDY version 10.00 \citep{ferland98} for nebulae around stars of different spectral types. We use the \citet{smith02} model grids at $Z=0.2$\Zsol~for main sequence O stars of spectral types O3-O6 and for WN and WC stars with core temperatures of 45,000 K-120,000 K; the WR core temperatures correspond to a stellar optical depth of 10. The grids use WM-Basic stellar atmosphere models \citep{pauldrach01} for the O stars and CMFGEN models \citep{hillier98} for the WR stars.

Since the ionization parameter is set by a combination of nebular geometry, density, and input radiation field (Equation~\ref{u_eqn}), we vary these parameters in our models. Our input stellar spectral energy distributions are scaled to a constant hydrogen-ionizing photon rate of $Q({\rm H})=10^{49.5}$ photons s$^{-1}$, which is the $Q({\rm H})$~value of an O3 V star at 0.2\Zsol~\citep{smith02}. We have run additional models with $Q({\rm H})=10^{50}$ photons s$^{-1}$; this change does not affect our results. For each spectral type, we vary the inner radius ($r_{\rm in}$)of the nebula from 0.5 \pc~to 8 \pc~and the hydrogen density from 0.01 to 1000 cm$^{-3}$. These inner radii correspond to a wide range of morphologies; $r_{\rm in}$ ranges from 0.2\% to 99.9\% of the outer radius of the ionized gas ($r_{\rm out}$). Morphologies with $r_{\rm in}$/$r_{\rm out} < 5\%$ only appear in models with $n_{\rm H} \leq 10$ cm$^{-3}$. Extremely thin shell morphologies, with $r_{\rm in}$/$r_{\rm out} > 95\%$, appear in some of the high density models and are marked with open circles in the figures. 

The gas phase metallicities match the stellar metallicities in each model run. We use element abundances from \citet{mcgaugh91}, interpolating between $Z=0.15$\Zsol~and $Z=0.30$\Zsol. For the abundances of all other elements, we use the CLOUDY gas-phase ISM abundances \citep{cowie86,savage96} and scale the abundances of the $\alpha$-elements to the O abundance from \citet{mcgaugh91}. We set the turbulence to 4 \kmps~and the grain abundances to 20\%~of the CLOUDY ISM value \citep{mathis77}. Changing the turbulence to 40 \kmps does not affect our results. 

\begin{figure*}
\epsscale{1.0}
\plotone{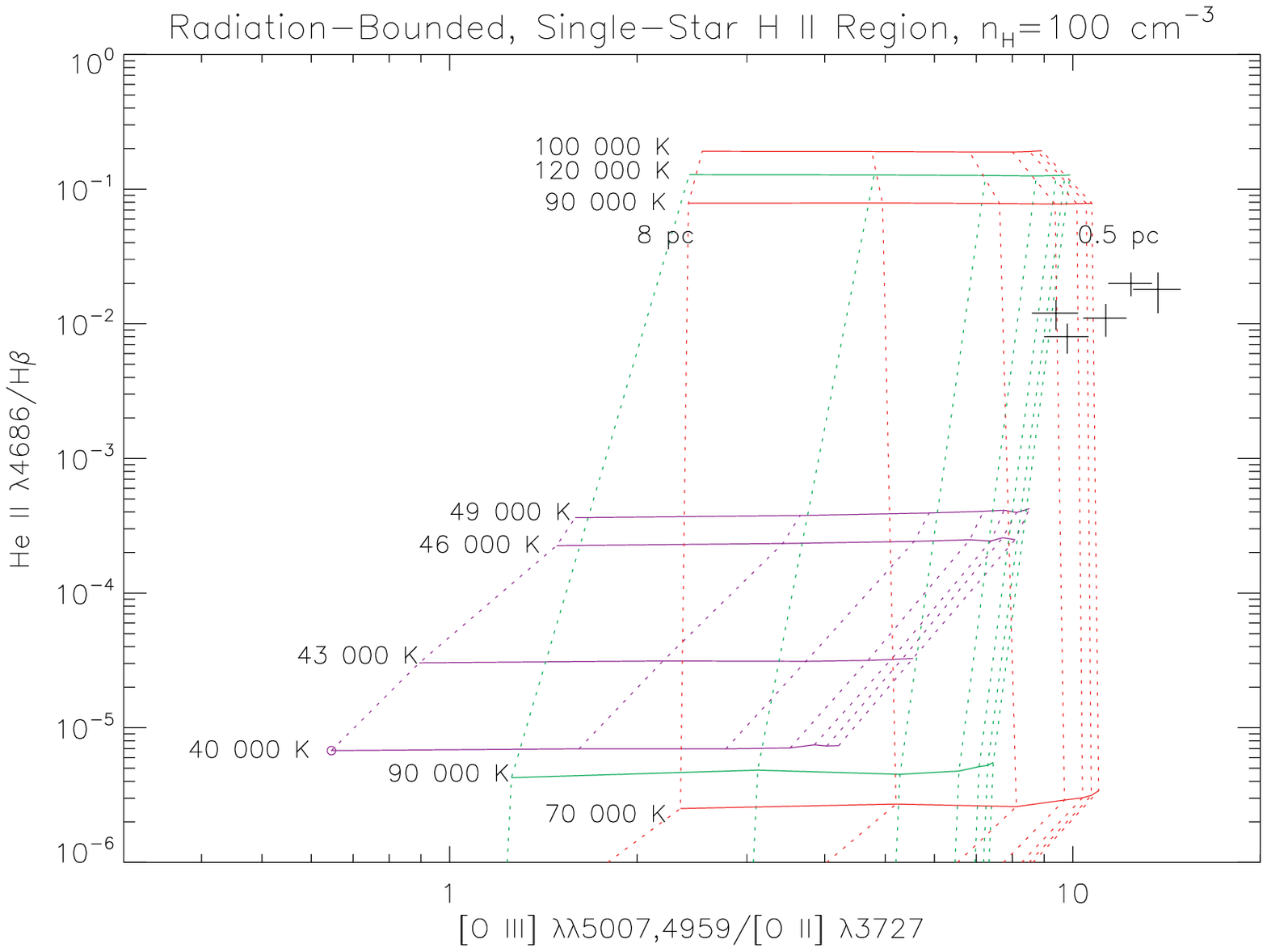}
\plotone{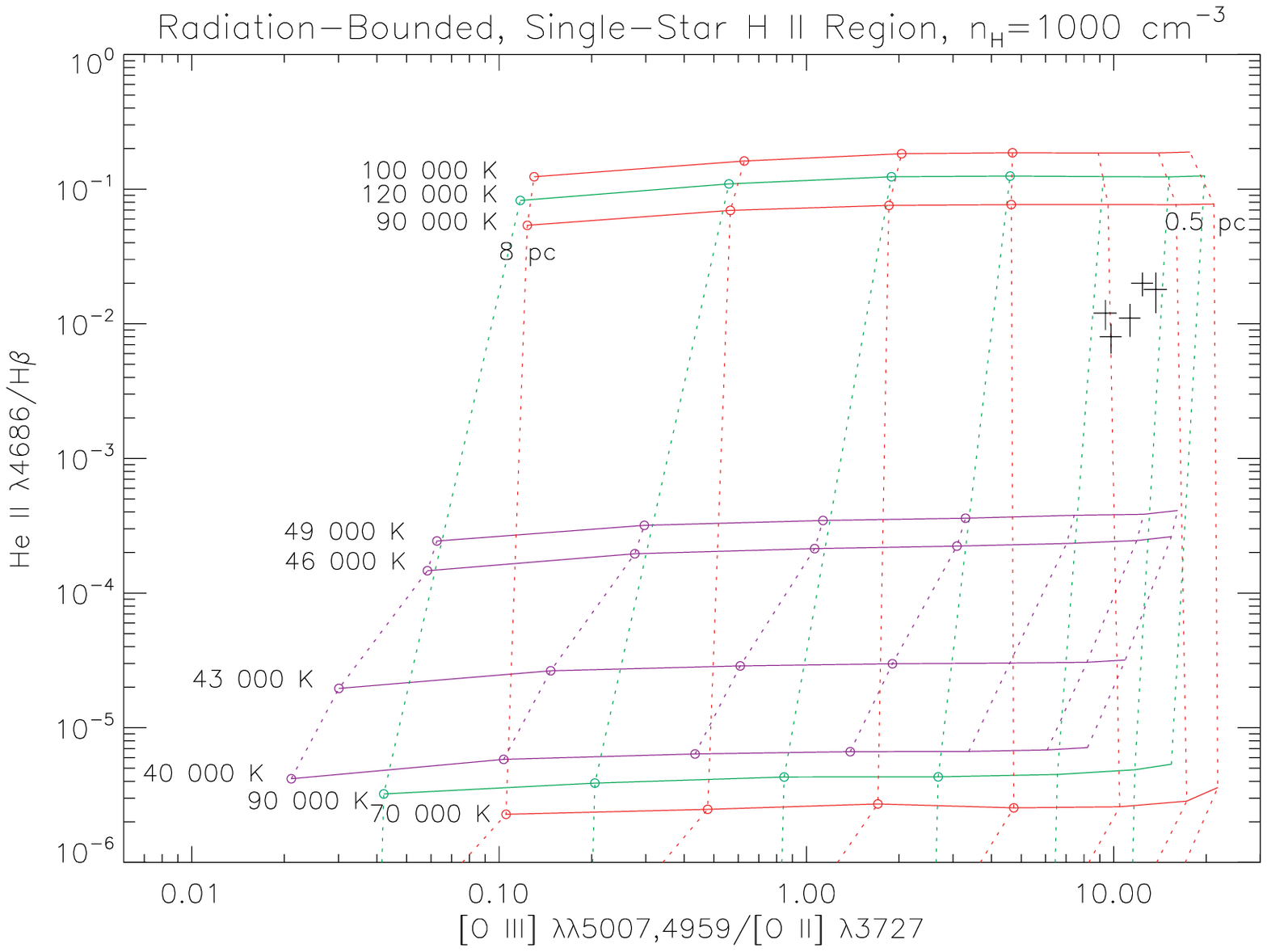}
\caption{Radiation-bounded (optically thick) CLOUDY photoionization model grids for single-star \hii~regions at densities of $n_{\rm H}=100$ cm$^{-3}$ (top) and $n_{\rm H}=1000$ cm$^{-3}$ (bottom). The black crosses show the positions of the five high-ionization Peas with \heii~detections and the 1$\sigma$ errors. The red grid corresponds to the WN star photoionization models, green corresponds to WC stars, and purple corresponds to main sequence O stars. Solid lines indicate constant stellar temperature; the O star effective temperatures and WR star core temperatures are labelled. Dotted lines indicate constant values of the inner nebular radius, from 0.5 pc to 8 pc. Thin shell models, where $r_{\rm in}$/$r_{\rm out} > 95\%$, are marked with open circles. Only the hottest WR stars match the observed line ratios.}
\label{fig_o_he2_rb}
\end{figure*}

We analyze the resulting line emission from both radiation-bounded and density-bounded scenarios by examining the cumulative line fluxes at different radii within the nebulae. The fluxes at low radii correspond to optically thin scenarios, since the gas has not fully absorbed all the ionizing photons at that radius. To evaluate the models, we compare the predicted and observed line ratios for \oii $\lambda$3727/H$\beta$, \neiii $\lambda$3869/H$\beta$, \heii $\lambda$4686/H$\beta$, \ariv $\lambda$4740/\ariii $\lambda$7135, \oiii $\lambda\lambda$5007,4959/\oii $\lambda$3727, and \hei $\lambda$5876/H$\beta$. We focus on \ariv~$\lambda$4740 rather than the stronger \ariv~$\lambda$4711 line due to contamination from \hei~$\lambda$4713. While \oiii/\oii~and \ariv/\ariii~are particularly sensitive to the ionization parameter or optical depth, \heii/H$\beta$ sets strong constraints on the temperature of the ionizing source. As shown in Figure~\ref{fig_o_he2_rb}, even though both WR and O star models can match the other line ratios, only the hottest, early-type WR stars, with core temperatures greater than 70,000 K, can produce the observed \heii/H$\beta$ ratios.

Using the He$^+$-ionizing photon rates from \citet{smith02} for the hottest WN and WC stars, we calculate the number of extreme early-type WR stars necessary to account for the observed \heii~luminosities. We use the helium emissivities and recombination coefficients from \citet{storey95} for $n_e=100$ cm$^{-3}$ and $T_e = 15,000$ K. Tables~\ref{table_wne} and \ref{table_wce} list the numbers of early-type WN (WNE) and WC (WCE) stars. The resulting WR/O ratios, calculated as in \S\ref{sec:stellar}, are a few percent, which matches the expected value for a 3-4 Myr old instantaneous burst. However, if the star formation is continuous, the Starburst99 models show that the WR/O ratios should never increase above 0.016 due to the continued production of new O stars. 

\begin{table*}
\vspace*{-0.2in}
\begin{center}
\caption{Constraints on WNE Stars in the Green Peas}
\label{table_wne}
{\scriptsize
\begin{tabular}{lcccc}
\hline
Galaxy & $N$(WNE stars)$^a$ & \multicolumn{2}{c}{$N$(WNE stars)$^b$} & WNE/O\\ 
& from \heii~$\lambda$4686 & \multicolumn{2}{c}{from \nv~$\lambda\lambda$4603-20} & from \heii~$\lambda$4686 \\
&  & $Z < 0.2$\Zsol & $Z \geq 0.2$\Zsol & \\
\hline 
\peaAs & 2100 & $\leq 200000$ & $\leq 30000$ & 0.02-0.05 \\
\peaBs & 10000 & $\leq 470000$ & $\leq 70000$ & 0.02-0.05 \\
\peaCs & 2600 & $\leq 270000$ & $\leq 41000$ & 0.02-0.03 \\
\peaDs & 2400 & $\leq 290000$ & $\leq 44000$ & 0.008-0.02 \\
\peaFs & 3000 & $\leq 240000$ & $\leq 36000$ & 0.009-0.03 \\
\hline
\end{tabular}
\flushleft{
$^a$Number of stars needed to produce the observed \heii~luminosity.\\
$^b$1$\sigma$ limit on the WNE population from \nv~$\lambda\lambda$4603-4620, using the WNE star \nv~luminosities of \citet{brinchmann08}.\\
}
\break
}
\end{center}
\end{table*}

\begin{table*}
\vspace*{-0.2in}
\begin{center}
\caption{Constraints on WCE Stars in the Green Peas}
\label{table_wce}
{\scriptsize
\begin{tabular}{lccc}
\hline
Galaxy & $N$(WCE stars)$^a$ &$N$(WCE stars)$^b$ & WCE/O\\ 
& from \heii~$\lambda$4686 & from \ciii+\civ~$\lambda\lambda$4650-4658 & from \heii~$\lambda$4686 \\
\hline
\peaAs & 1100 & ... & 0.009-0.02\\
\peaBs & 5200 & ... & 0.01-0.03 \\
\peaCs & 1300 & 1200 & 0.01\\
\peaDs & 1200 & 1900 & 0.004-0.01\\
\peaFs &  1500 & 1400 & 0.004-0.02\\
\hline
\end{tabular}
\flushleft{
$^a$Number of stars needed to produce the observed \heii~luminosity.\\
$^b$Inferred WCE population from \ciii+\civ~$\lambda\lambda$4650-4658, using the WCE star \ciii~and \civ~luminosities of \citet{crowther06}. The \ciii+\civ~emission may be contaminated by nebular \feiii.\\
}
\break
}
\end{center}
\end{table*}

While the WR/O ratios are reasonable, at least for an instantaneous burst, we also consider whether the constraints on WR stellar features are consistent with this number of WR stars. The blue bump emission of WNE stars comes from \nv~$\lambda$4603-4620, and the WNE stars have weaker blue bump luminosities than the WNL stars \citep{brinchmann08}. Thus, the required population of 2000-9000 WNE stars could easily go unnoticed in the Peas' spectra (see Table~\ref{table_wne}). WNE stars do have strong and broad \heii~$\lambda$4686 emission, however. Adopting the $Z < 0.2$\Zsol and $Z \geq 0.2$\Zsol~WNE $\lambda$4686 luminosities of \citet{brinchmann08}, we find that the broad component should be 6\%-30\% the strength of the nebular component, respectively. Although we do not see any broad \heii~component in the stacked spectra, we cannot rule out the presence of WNE stars, particularly if the lower metallicity luminosities are valid. 

The WCE stars should also have broad \heii~emission, blended with \ciii~$\lambda$4650 and \civ~$\lambda$4658 \citep[e.g.,][]{crowther06}. From the WCE $\lambda$4686 and $\lambda$4650 luminosities of \citet{crowther06}, the broad \heii~component should be 89\% as strong as the nebular emission. The lack of this broad emission in the stacked spectra (Figure~\ref{fig_stack}) indicates that thousands of WCE stars are unlikely to be present in most Peas. One Pea that might plausibly have WCE stars is \peaFs, which appears to have broad $\lambda\lambda$4650-4670 emission (Figure~\ref{fig_bb}). The strength of this emission is consistent with the number of WCE stars required to produce the observed \heii. As with the WNE stars, Starburst99 models can match the inferred WC/O ratios with an instantaneous burst but not with continuous star formation. 

The above calculations show that photoionization by early-type WR stars is possible. In reality, the nebular gas in the Peas is likely photoionized by stars of many different spectral types. Consequently, we ran CLOUDY photoionization models with the $10^6$\Msol~Starburst99 instantaneous burst model (\S\ref{sec:age}) as the input ionizing spectrum. The nebular model parameters are the same as described for the single-star models, except for the inner radius, which we varied from 20-81 \pc.  As before, these models occasionally produce thin shell nebulae, where $r_{\rm in}$/$r_{\rm out} > 95\%$. We denote radiation-bounded, thin shell models by open circles in the figures. The other optically thick models have $r_{\rm in}$/$r_{\rm out} $ between 8.6 and 95\%. The thin shell models may be valid if dense filaments dominate the Peas' nebular emission. In addition, unlike the single-star CLOUDY models, the hydrogen-ionizing photon rate, $Q({\rm H})$, varies as a function of the age of the starburst.

\begin{figure*}
\epsscale{1.0}
\plotone{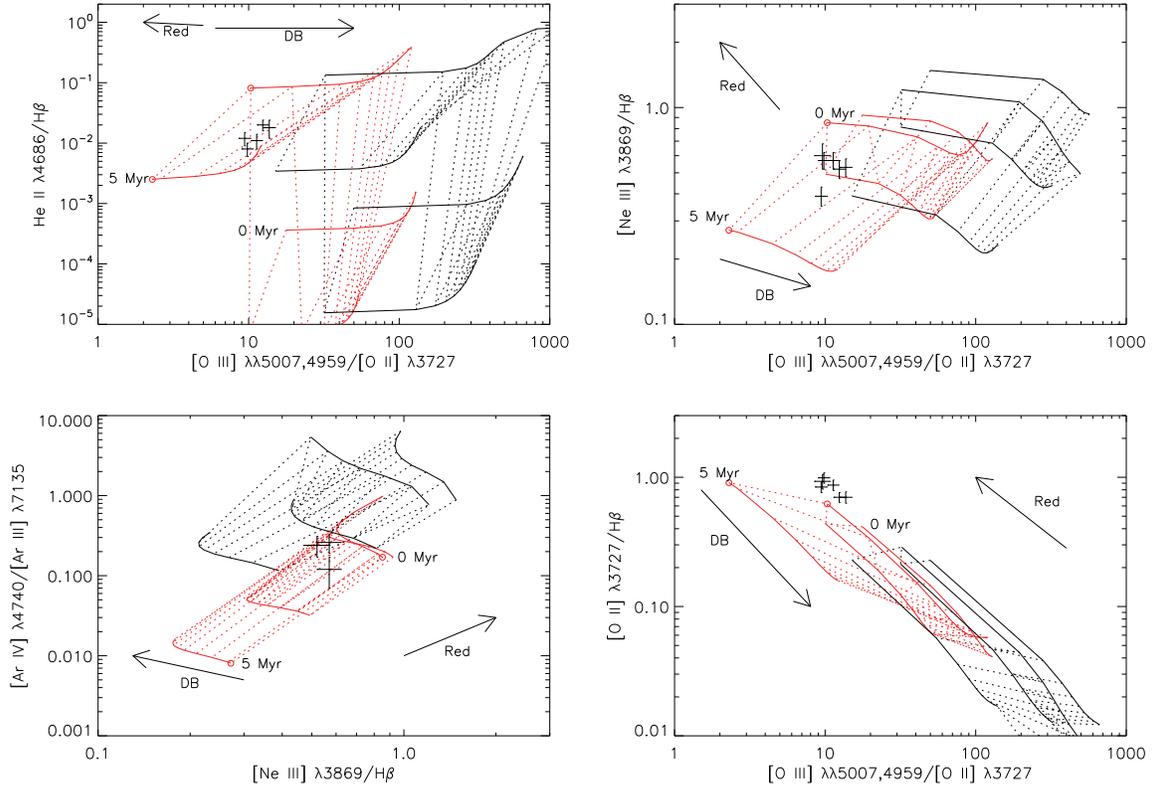}
\caption{CLOUDY grids with a 10$^6$\Msol~ionizing cluster and a nebular density of $n_{\rm H}$ = 100 cm$^{-3}$. The high-ionization Peas are marked with black crosses indicating the 1$\sigma$ errors for the line ratios. The black grids have an inner radius of 20 pc, while the red grids have an inner radius of 81 pc. Solid lines indicate the age of the cluster and correspond to ages of 0, 3, 4, and 5 Myr. The arrow labeled `DB' indicates the direction of decreasing optical depth (increasing density-bounding). Constant values of the optical depth are indicated by the dashed lines in each grid. The dashed lines show the emission as the nebular size is decreased from its full ionized extent to 20\% of its original thickness, corresponding to a decrease in the optical depth.  The arrow labeled `Red' indicates the reddening vector; the plotted points would move parallel to this vector as the reddening correction is increased. Open circles mark thin shell models, where $r_{\rm in}$/$r_{\rm out} > 95\%$ and $r_{\rm out}$ is the outer radius of nebula for the radiation-bounded case.}
\label{fig_sb99n2}
\end{figure*}

\begin{figure*}
\epsscale{1.0}
\plotone{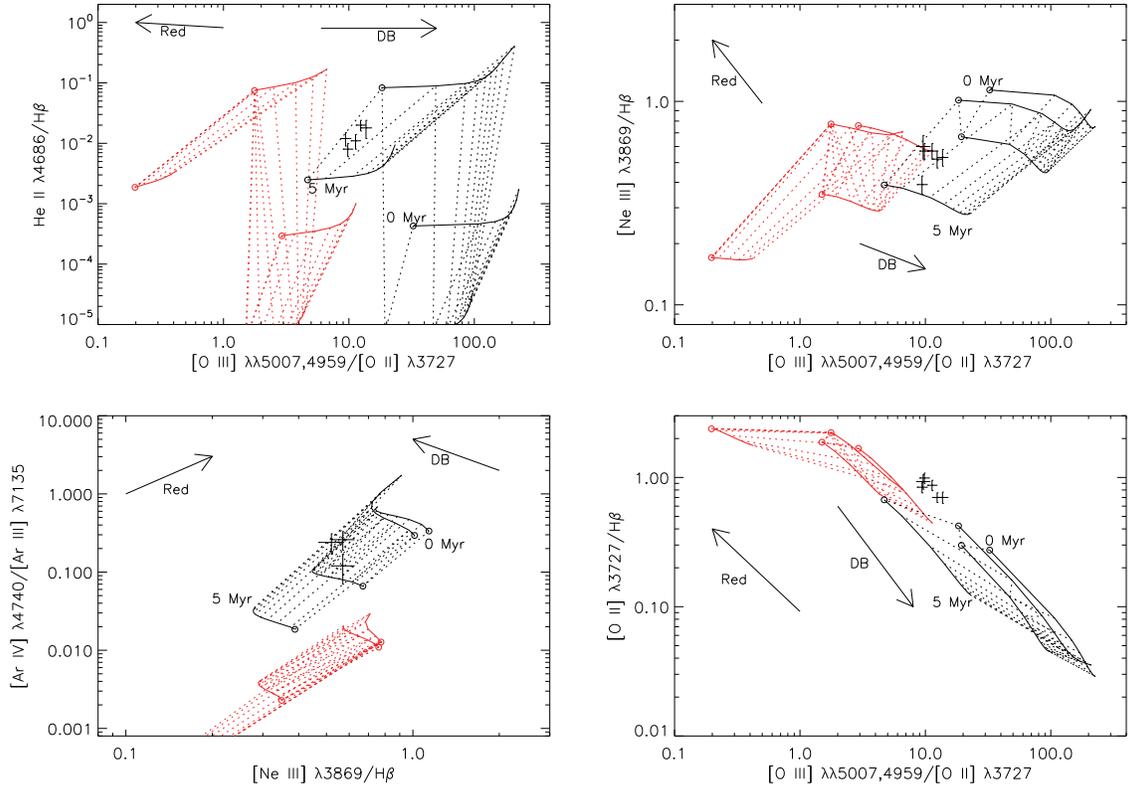}
\caption{CLOUDY and Starburst99 grids for $n_{\rm H}$ = 1000 cm$^{-3}$. Colors and symbols are described in Figure~\ref{fig_sb99n2}.}
\label{fig_sb99n3}
\end{figure*}

\begin{figure*}
\epsscale{1.0}
\plotone{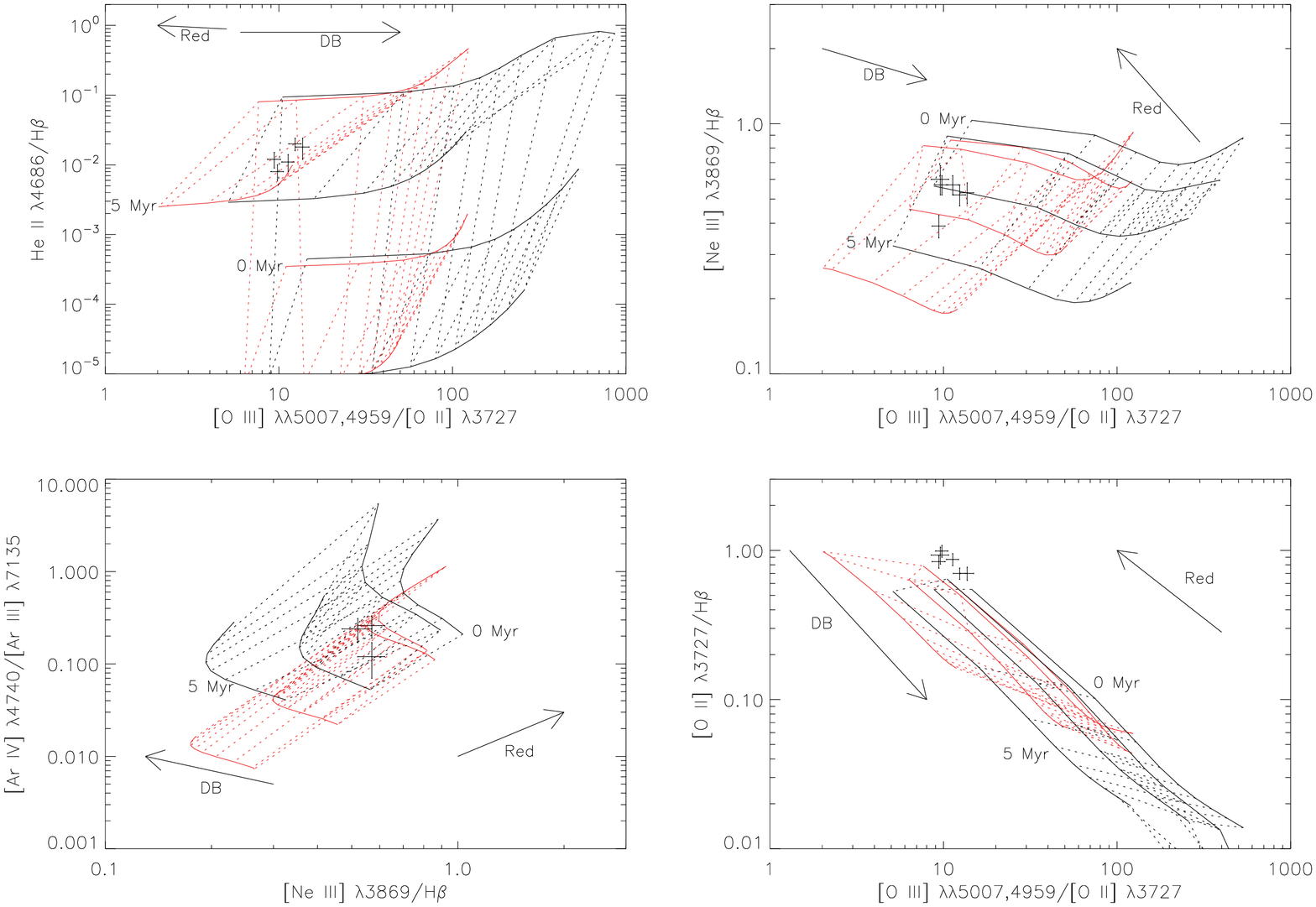}
\caption{CLOUDY and Starburst99 grids for $n_{\rm H}$ = 100 cm$^{-3}$ and a filling factor of 0.1. Colors and symbols are described in Figure~\ref{fig_sb99n2}.}
\label{fig_sb99ff}
\end{figure*}

\begin{figure*}
\epsscale{1.0}
\plotone{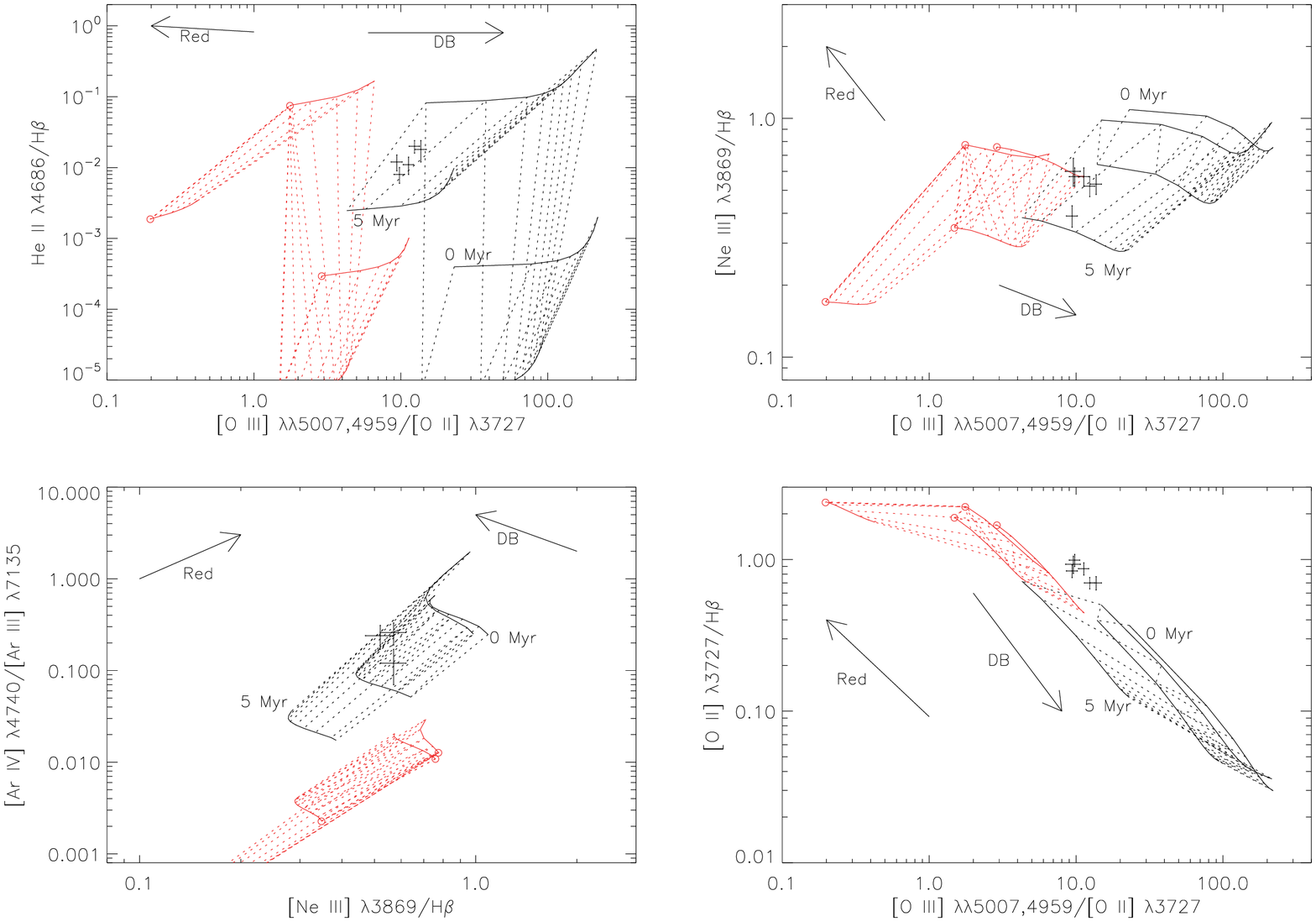}
\caption{CLOUDY and Starburst99 grids for $n_{\rm H}$ = 1000 cm$^{-3}$ and a filling factor of 0.1. Colors and symbols are described in Figure~\ref{fig_sb99n2}.}
\label{fig_sb99n3ff}
\end{figure*}

We plot the CLOUDY and Starburst99 results for several emission line ratios in Figures~\ref{fig_sb99n2}-\ref{fig_sb99n3ff}. Ages of $\sim4$-5 Myr, when the burst contains a high fraction of WR stars, do the best job of fitting the observed \heii/H$\beta$~ratios. At all ages, the models have difficulty matching the observed \oii/H$\beta$ ratios. Part of this discrepancy arises from the modeled nebular morphology. The models treat the ionized gas as a shell surrounding an inner cavity. While this is an appropriate model for a single superbubble, the ionized gas in the Peas is likely much less homogeneous. By adding a filling factor and treating the ionized gas as dense clumps in a lower density medium, we obtain an improved fit to the \oii/H$\beta$ emission (Figures~\ref{fig_sb99ff} and \ref{fig_sb99n3ff}; see also Zastrow et al. 2013). We adopt filling factors of 1, 0.1, and 0.01, where the filling factor indicates the volume of the gas occupied by clumps of a given density. The remaining volume is treated as a vacuum. The filling factor models are still offset to lower \oiii/\oii~ratios than we observe, which may indicate a need for higher temperature sources. As we discuss in the next section, unusually hot massive stars could explain this discrepancy.

The CLOUDY models support both optically thick and optically thin scenarios, depending on the nebular parameters. In general, the models with higher densities and larger inner radii imply lower optical depths. For densities $n < 1000$ cm$^{-3}$, the strength of the \oii/H$\beta$~ratio implies a high optical depth, while the combination of \neiii/H$\beta$~and \ariv/\ariii~ratios suggests a low optical depth. For the $n = 1000$ cm$^{-3}$ models, applicable to Pea \peaBs, the \oii/H$\beta$ ratio supports a low optical depth if the inner radius is larger than 50 pc or if the filling factor is 0.01. The other line ratios generally favor low optical depths for $n = 1000$ cm$^{-3}$ and an inner radius larger than 30 pc and high optical depths in the other models. We discuss the optical depth further in \S\ref{sec:discussion}. 

Photoionization by WR stars is a viable explanation for the Peas' line ratios, as long as the starbursts are old enough to have a sizeable  WR population. The clear detection of the \hei~$\lambda$3819 line in Peas \peaBs, \peaCs, and \peaDs~(see \S\ref{sec:age}) calls this interpretation into question. In addition, the WR hypothesis depends strongly on our choice of WR line luminosities. While hot WR stars may account for part of the \heii~emission in the high-ionization Peas, additional causes of the emission may be required for the youngest Peas. 

\subsubsection{Homogeneously-Evolved O Stars}
\label{sec:nebular:hoto}
If WR stars are not present, hot, low-metallicity O stars could perhaps be the source of the hard ionizing photons. In the past few decades, studies have increasingly recognized that massive, rotating stars at low metallicity can achieve much higher effective temperatures than ordinary O stars. At low metallicity, weaker stellar winds make it difficult for stars to lose angular momentum \citep{langer98}, and as a result, stars can maintain high rotational velocities. The associated rotational mixing may then result in chemically homogeneous or quasi-homogeneous evolution \citep[e.g.,][]{maeder87,yoon06}. Chemically homogeneous O stars may have effective temperatures much higher than expected for their mass. For instance, \citet{brott11} show that, at the metallicity of the LMC, stars more massive than 20\Msol~may reach temperatures as high as 60,000-70,000 K. The highest mass stars are the most likely to evolve homogeneously and reach such high temperatures \citep[e.g.,][]{brott11}. These stars also evolve more quickly than lower mass O stars and could be present in young starbursts. For instance, \citet{brott11} show that 60 \Msol~stars at Small Magellanic Cloud (SMC) metallicity can reach 60,000 K within 3.4 Myr. These high temperature O stars may be plentiful in low-metallicity starbursts and could provide the necessary high-energy photons for the Peas' \heii~emission.

\begin{figure*}
\epsscale{1.0}
\plotone{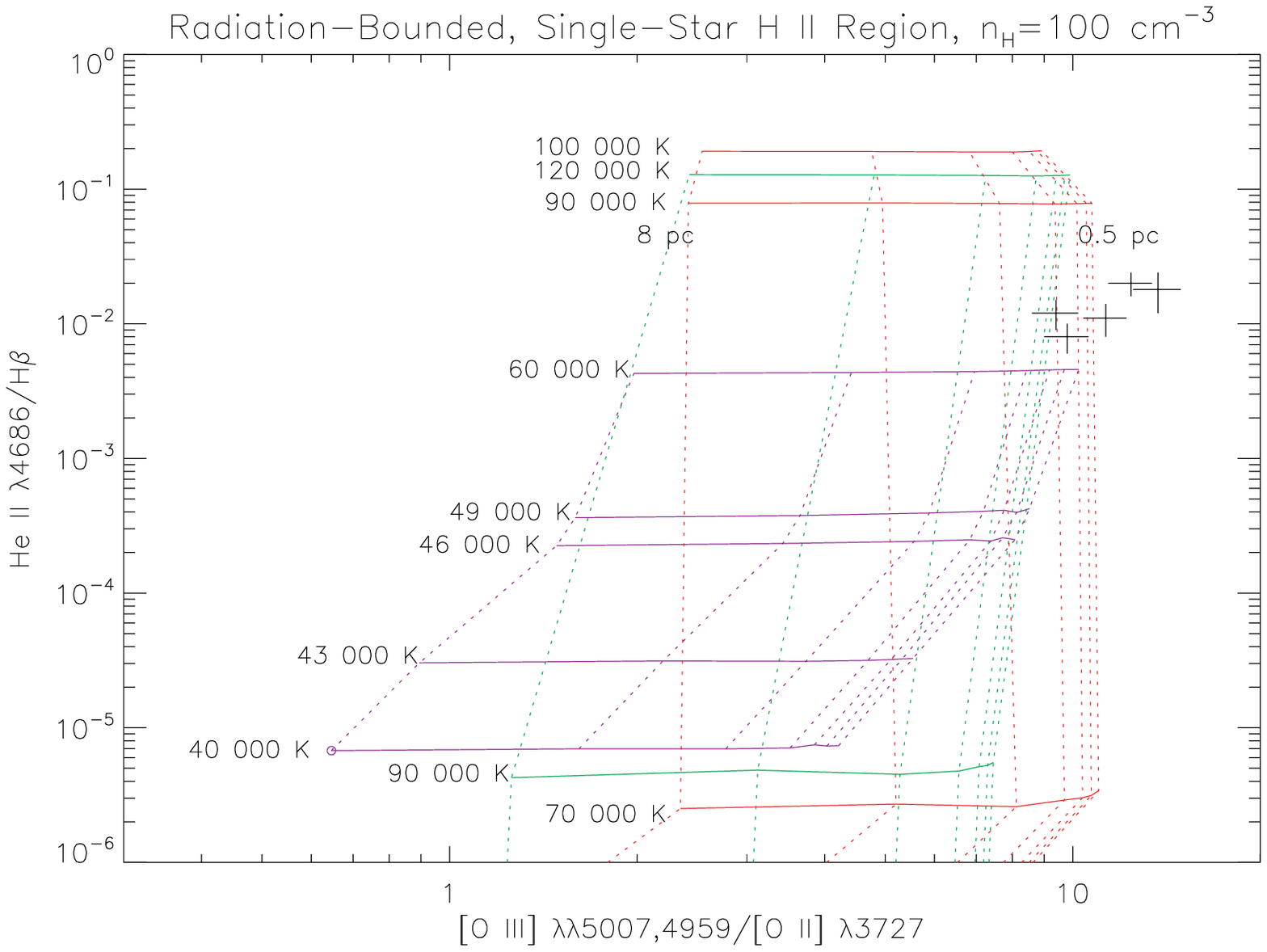}
\plotone{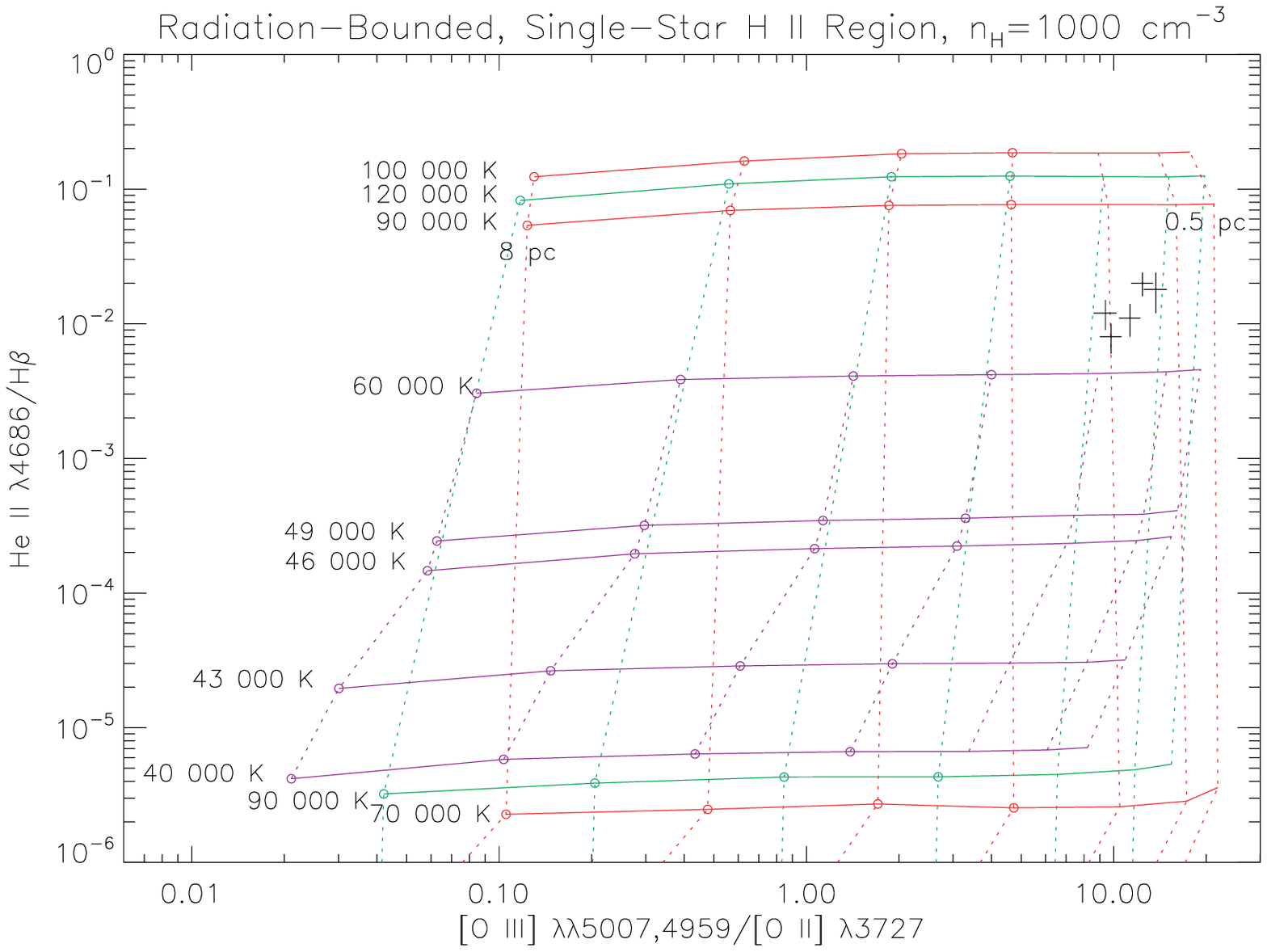}
\caption{As in Figure~\ref{fig_o_he2_rb}, with the inclusion of the 60,000 K O star model of \citet{kudritzki02}.}
\label{fig_o60}
\end{figure*}

As a test, we consider the 60,000 K O star models of \citet{kudritzki02}. We ran a single-star \hii~region CLOUDY model as described in \S\ref{sec:nebular:wr} for the $T_{\rm eff} = 60,000$ K, $\log(L/\Lsol)=7.03$ luminosity O star model. Figure~\ref{fig_o60} shows that a 60,000 K O star continues the trend of increasing \heii/H$\beta$ with effective temperature. The 60,000 K O star model still under-predicts the observed \heii/H$\beta$~ratio. However, O stars with temperatures above 60,000 K could result in higher \heii/H$\beta$~ratios. 

To determine the number of hot O stars needed to generate the observed \heii~luminosities, we use the ionizing photon rates from \citet{kudritzki02} for the $\log(L/\Lsol)=7.03$, 60,000 K star \citep[see Figure 11 of][]{kudritzki02} and emissivities and recombination rates for $T_e = 15,000$ K and $n_{\rm H} = 100$ cm$^{-3}$ \citep{storey95}. Table~\ref{table_o60} lists the calculated numbers of hot O stars. The number of 60,000 K O stars required to explain the Peas' emission represent 0.4-3\% of their total O star populations. For comparison, at ages less than 3 Myr, the instantaneous burst Starburst99 model shows that 8-18\%~of O stars have spectral types earlier than O4. Therefore, if the earliest spectral types evolve homogeneously, this population could account for the high ionization lines observed in the Peas. Nevertheless, we caution that the existence of chemically homogeneous massive stars has not yet been confirmed, even in local, low-metallicity galaxies \citep[e.g.,][]{brott11b}. More research is necessary to establish whether homogeneous stars are present in the Peas.

\begin{table*}
\vspace*{-0.2in}
\begin{center}
\caption{Hot O Stars in the Green Peas}
\label{table_o60}
{\scriptsize
\begin{tabular}{lcc}
\hline
Galaxy & $N$(60,000 K O stars) & 60,000 K O Stars / Total O Stars \\
& from \heii~$\lambda$4686 & \\ 
\hline
\peaAs & 1200 & 0.01-0.03\\
\peaBs & 5800 & 0.01-0.03 \\
\peaCs & 1500 & 0.01-0.02\\
\peaDs & 1400 & 0.004-0.01\\
\peaFs &  1700 & 0.005-0.02\\
\hline
\end{tabular}
}
\end{center}
\end{table*}

\subsection{Active Galactic Nuclei}
\label{sec:agn}

\begin{figure*}
\epsscale{1.0}
\plotone{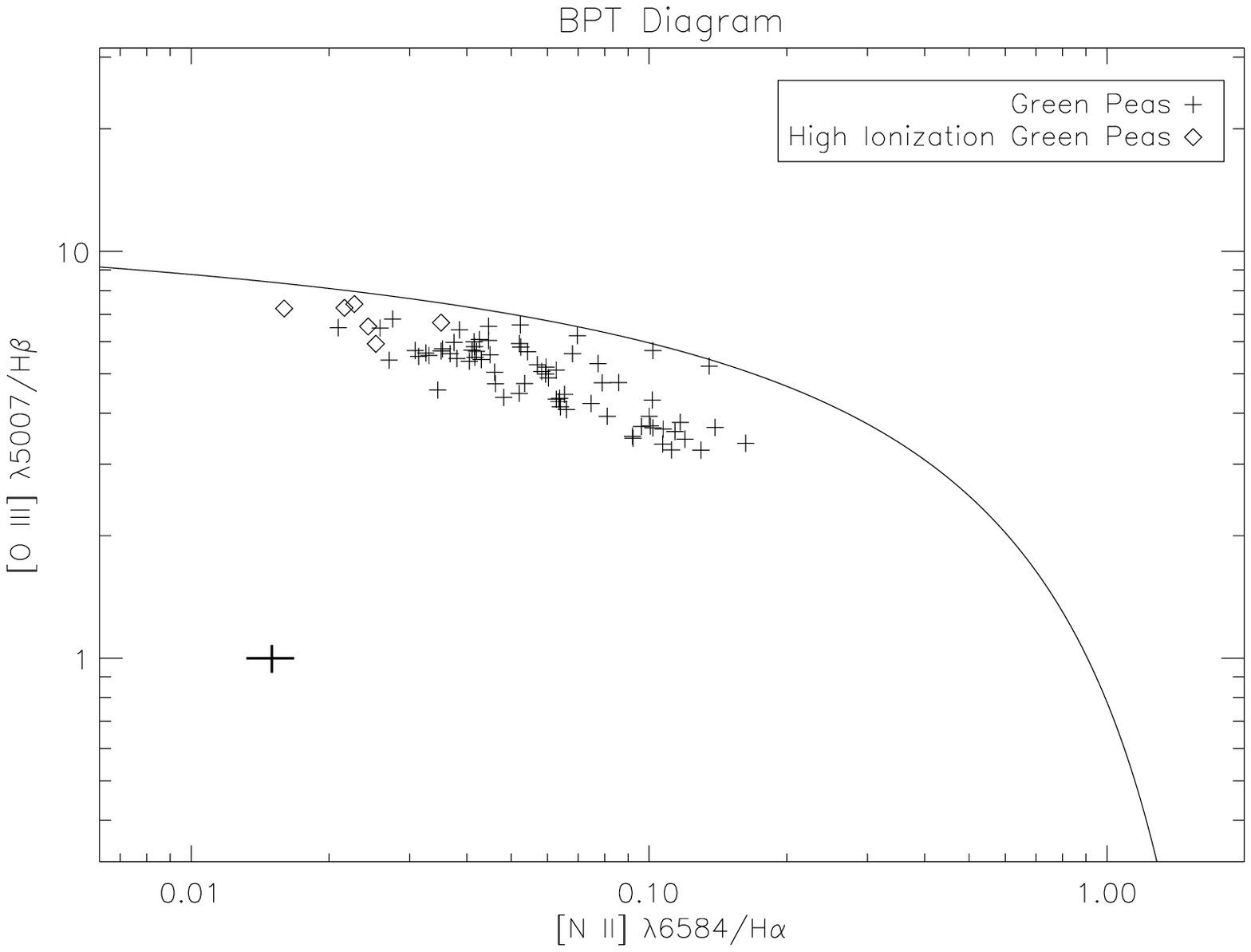}
\plotone{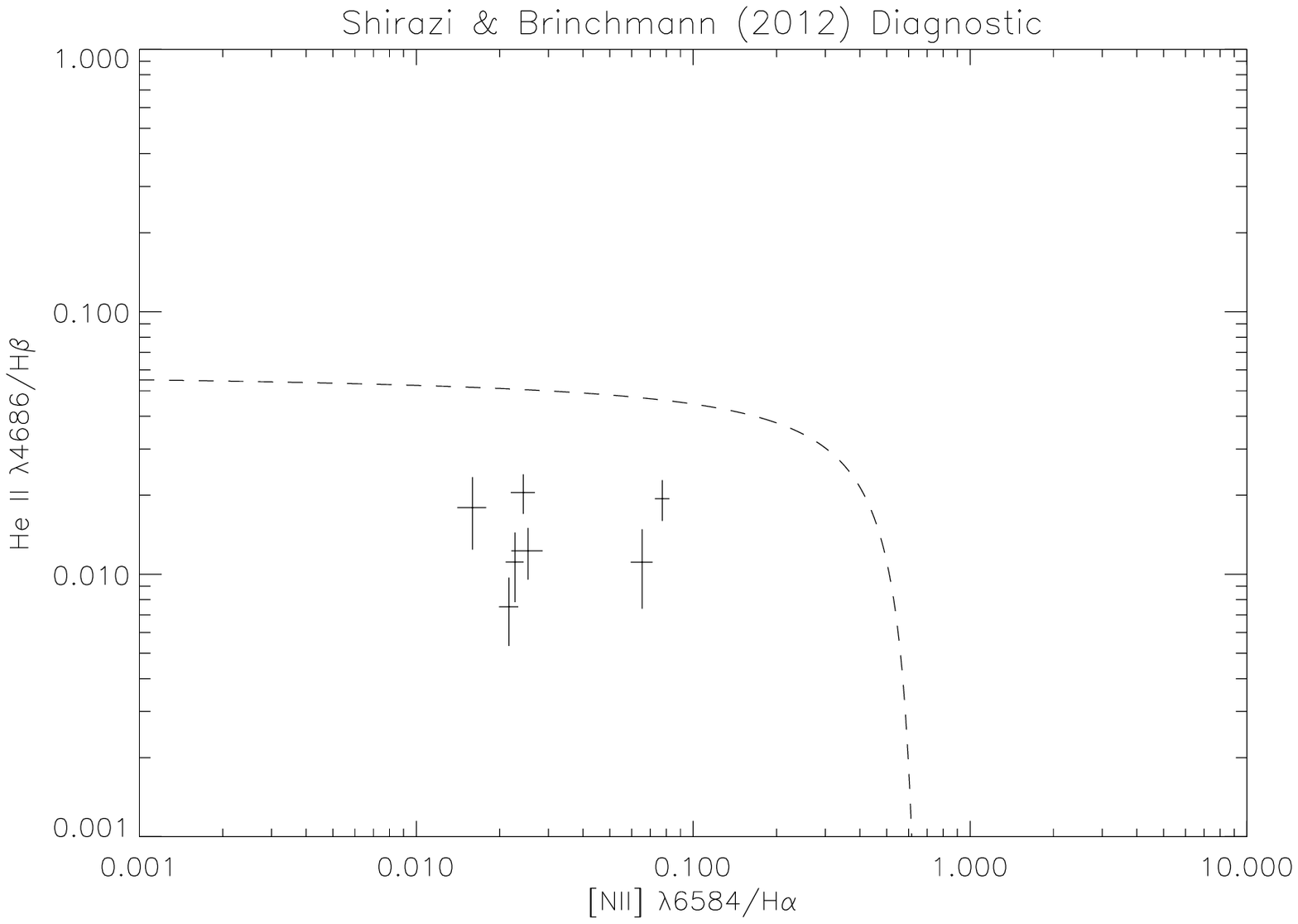}
\caption{The top panel shows the Green Peas' positions on the BPT diagram \citep{baldwin81}. The solid line indicates the \citet{kewley01} extreme starburst line. The Peas lie close to the dividing line between starbursts and AGN. Representative 1$\sigma$~errors for the extreme Peas are shown by the black cross in the lower left. The bottom panel shows the AGN diagnostic of \citet{shirazi12} for the Peas with \heii~$\lambda$4686 detections; the line indicates the point at which an AGN contributes 10\% of the \heii~emission. In this diagram, the Peas are clearly separate from the AGN regime. }
\label{fig_agn}
\end{figure*}

Although the BPT diagram indicates the Peas are star-forming, many of the galaxies lie close to the maximum starburst line of \citet{kewley01}, which marks a theoretical separation between starbursts and AGN. Given their proximity to this line, the Peas could have some AGN contribution to their emission (Figure~\ref{fig_agn}). \citet{shirazi12} show that a plot of \heii~$\lambda$4686/H$\beta$ versus \nii~$\lambda$6584/H$\alpha$ results in a much cleaner separation between AGN and star-forming galaxies. Figure~\ref{fig_agn} shows the \citet{shirazi12} diagnostic for the Green Peas with \heii~$\lambda$4686 detections. The line indicates the region where AGN contribute 10\% of the observed \heii~flux. All the observed \heii~fluxes lie below this line, solidly in the star-forming regime, confirming that AGN activity contributes negligibly to the high ionization emission in the Peas. 

\subsection{High-Mass X-ray Binaries}
\label{sec:hmxbs}

Following a starburst event, HMXBs could be a significant source of hard photons. In addition, HMXB luminosities may be higher at the low metallicities characteristic of the Peas. Given the increase of X-ray heating with metallicity, black holes may be able to accrete more gas at low metallicity \citep{thuan04}. Alternatively, the weaker winds and lower mass loss rates of low metallicity stars may lead to the formation of higher mass black holes \citep{thuan05,fragos12}. In three low-metallicity blue compact dwarf (BCD) galaxies, \citet{kaaret11} find X-ray luminosities an order of magnitude higher than expected from the SFR. The metallicities of these sources are lower than the Green Peas, however, with $Z < 0.07$\Zsol; HMXB luminosities at 0.2\Zsol~might not show such a dramatic increase.

While an HMXB scenario is appealing, it faces a number of problems when applied to the Peas. The Peas' young ages may be at odds with the HMXB interpretation, as the formation of HMXBs requires both the formation of a compact object and additional time for the evolution of the companion star \citep{ghosh01}. The first HMXBs appear 4 Myr after a burst of star formation \citep{linden10}, which is inconsistent with the upper age limits for at least one of the Peas. In addition, the simulations of \citet{linden10} show that HMXB numbers and luminosities have a complex dependence on metallicity. In fact, at 5 Myr after a burst, high-metallicity galaxies have a greater number of luminous HMXBs than low-metallicity galaxies. 

At a burst age of 4-5 Myr and $Z=0.2$\Zsol, \citet{linden10} predict the formation of 5 HMXBs with X-ray luminosities greater than $10^{36}$\ergps per $10^6$\Msol~burst. Four of these HMXBs would qualify as ultra-luminous X-ray sources (ULXs) with X-ray luminosities above $10^{39}$\ergps. The ULX population peaks at this age, while the HMXB population reaches its peak a few Myr later \citep{linden10}. Scaling these numbers to the calculated burst masses of the Peas, we would expect 68-854 luminous HMXBs and 54-683 ULXs at an age of 4-5 Myr. \citet{kaaret04} report the detection of strong \heii~$\lambda$4686 emission, with a luminosity of $2.7 \times 10^{36}$\ergps, around a ULX nebula in Holmberg II; the galaxy has a metallicity near 0.2\Zsol \citep{moustakas10}. Adopting this luminosity for the \heii~emission produced by a ULX, the extreme Peas would need 2100-11,000 ULXs to account for their \heii~luminosities. Even if the Peas are old enough to have an HMXB population, this number of ULXs is more than an order of magnitude larger than expected. The \citet{grimm03} relation between the SFR and the number of luminous HMXBs results in an even worse discrepancy, with only 20-89 HMXBs expected in the Peas. We conclude that HMXBs are an unlikely source of the observed \heii~emission.

\subsection{Shocks}
\label{sec:shocks}

\citet{guseva00} and \citet{thuan05} propose fast, radiative shocks as a means of generating \heii~emission in galaxies without WR stars. In a study of BCDs, \citet{thuan05} observe emission from Ne$^{4+}$~and Fe$^{4+}$, which have ionization potentials even higher than He$^{+}$. Radiative shocks can naturally account for the existence of these highly ionized emission lines. In addition, the \oiii~$\lambda\lambda$4959,5007 and H$\beta$~lines in the BCDs exhibit broad wings indicative of high velocity gas \citep{thuan05}. Observations of the BCD SBS 0335-052E show that the \heii~and \nev~emission originate in a separate region from the \ariv~and \oiii~emission \citep{izotov06}. The latter emission lines are likely photoionized by the youngest clusters in the galaxy. The \heii~emission, on the other hand, is associated with older clusters, which supports an interpretation of supernova remnant (SNR) shock ionization \citep{izotov06}. \citet{thuan05} argue that fast, radiative shocks are most effective in dense, low-metallicity, super star clusters, conditions which likely characterize the star-forming environments of the Green Peas. 

Supernovae, galactic outflows, or mergers could lead to fast shocks \citep{shirazi12}, and the Peas show evidence for these processes. Broad emission-line wings and multiple velocity components indicate high-velocity inflows and outflows in the Peas \citep{amorin12b}. The Peas' offset in the mass-metallicity relation likewise suggests the presence of low-metallicity gas inflows or metallicity-dependent, supernova-driven outflows \citep{amorin10}. Merger-induced shocks may also be present, as shown by the disturbed morphologies and presence of companion galaxies in {\it HST} images of the Peas \citep{overzier08, cardamone09}. 

\begin{figure*}
\epsscale{1.0}
\plotone{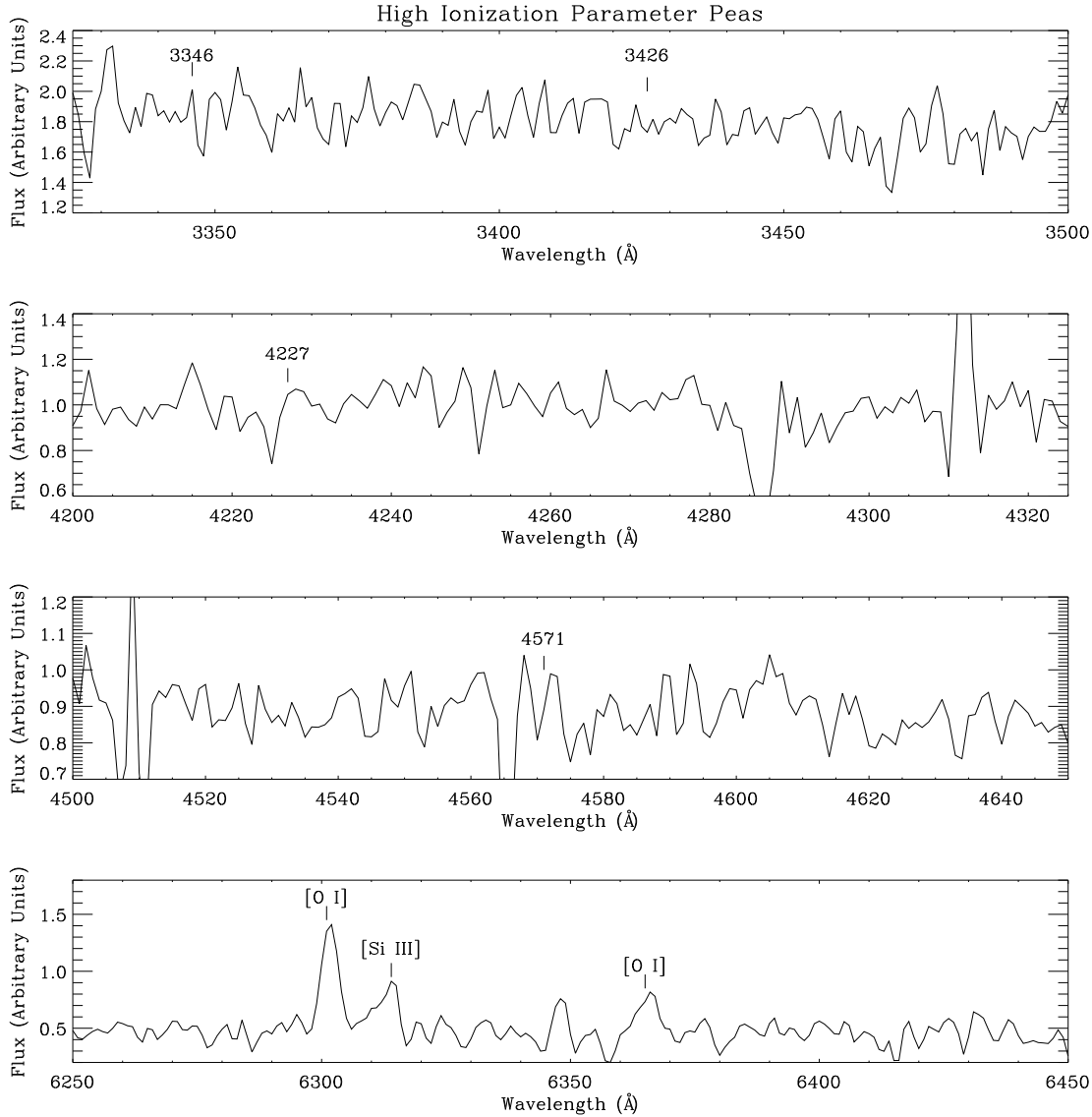}
\caption{The panels show the stacked spectrum of the high \oiii/\oii~ratio Peas for the spectral regions near shock lines. \nev~$\lambda$3346, \nev~$\lambda$3426, \fev~$\lambda$4227, and \mgi~$\lambda$4571 are not detected. The positions of these wavelengths are indicated. \oi~$\lambda$6300 is shown in the bottom panel.}
\label{fig_shocknd}
\end{figure*}

We detect emission from \oi~$\lambda$6300 and \none~$\lambda$5200, which are characteristic of shocks. Weak \none~emission is visible in Pea \peaCs's spectrum as well as in the stacked spectra. We do not detect the \nev~or \fev~lines discussed by \citet{thuan05}, but their data show that these lines are substantially weaker than \heii. Another non-detected shock line is \mgi~$\lambda$4571. The Mappings III shock models of \citet{allen08} show that low ratios of \mgi/\heii~can only be obtained for high magnetic field strengths, $\gtrsim$10$\mu$G. Magnetic field strengths of 20-50$\mu$G have been observed in starburst galaxies \citep[e.g.,][]{klein88, chyzy04}. Radio observations indicate that the Green Peas have similar magnetic field strengths of $> 30 \mu$G \citep{chakraborti12}, consistent with the non-detection of \mgi. Figure~\ref{fig_shocknd} shows the detected and non-detected shock lines in the high-ionization Pea stacked spectrum.

From the Mappings III shock model grids \citep{allen08}, the observed \heii/\oi~ratios in the extreme Peas support shock velocities between $\sim200$~and $\sim700$~\kmps. The observed H$\alpha$ full-width-at-zero-intensity values show that the Peas do have gas moving at these velocities. In the extreme Peas, these broad wings range from $\sim\pm$400-800 \kmps, similar to the velocities observed by \citet{thuan05}. We do not detect any significant velocity shift in the centers of the \heii~or \oi~lines, and their FWHM values are similar to the H$\alpha$ FWHM. Higher resolution observations are necessary to determine whether the \heii~emission has any unusual velocity components. 

\begin{figure*}
\epsscale{1.0}
\plotone{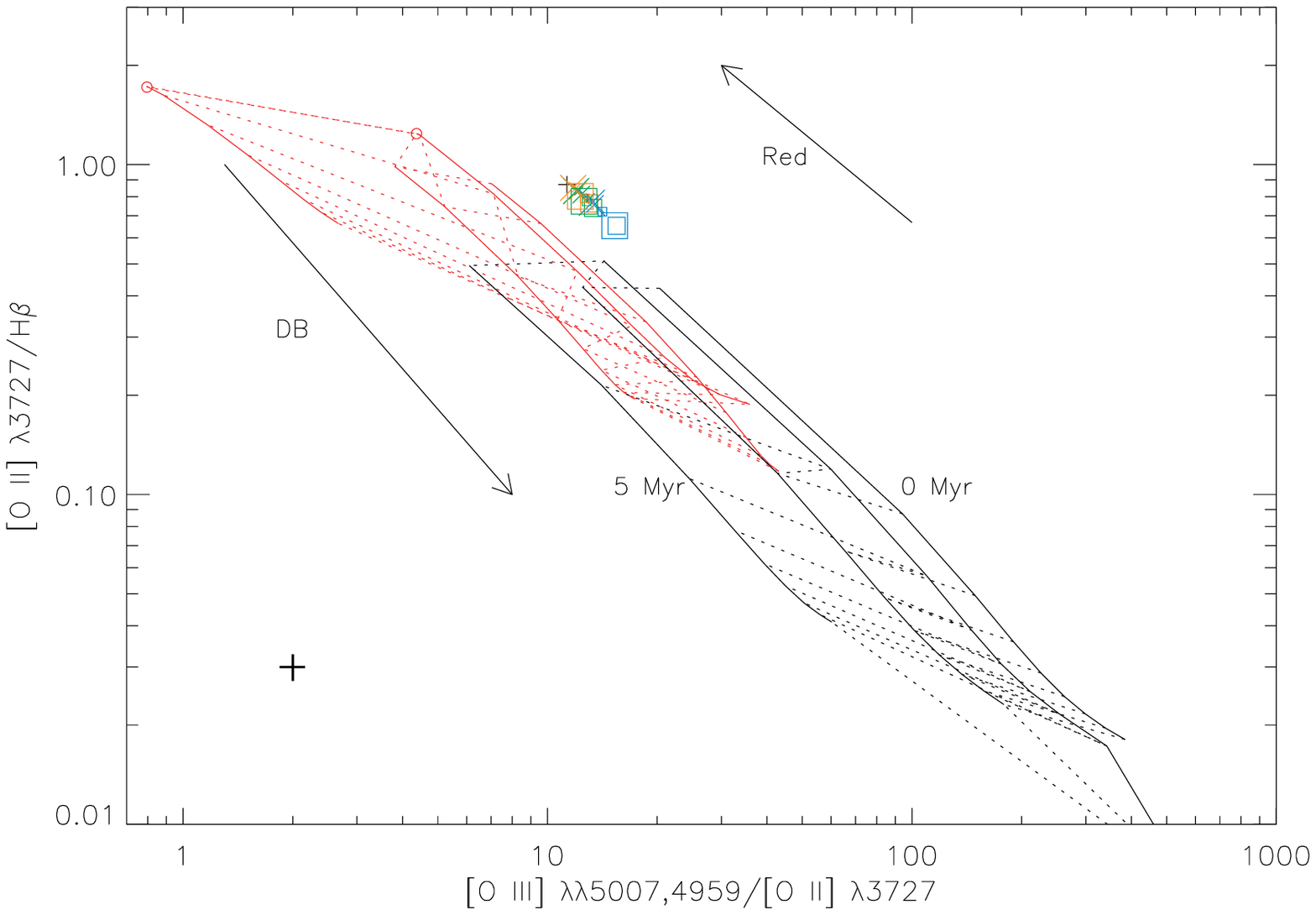}
\caption{The grid is described in Figure~\ref{fig_sb99n2}. The figure shows the CLOUDY and Starburst99 grids for $n_{\rm H}$ = 300 cm$^{-3}$ and a filling factor of 0.1. The black cross indicates the observed ratios of Pea \peaCs. The 1$\sigma$~errors are indicated by the cross in the lower left of the plot. The squares and X's indicate the line ratios with the shock contribution subtracted. The squares correspond to shock models at LMC metallicity, and the X's indicate SMC metallicity. The symbol size indicates the magnetic field strength; large symbols correspond to 10 $\mu$G models, intermediate-size symbols correspond to 3.23 $\mu$G, and small symbols correspond to 0.5 $\mu$G. Symbol colors denote shock velocity: blue for 300 \kmps~models, green for 500 \kmps~models, and orange for 1000 \kmps~models. Accounting for a shock contribution to the Peas' emission decreases their inferred optical depth.}
\label{fig_shock50}
\end{figure*}

\begin{figure*}
\epsscale{1.0}
\plotone{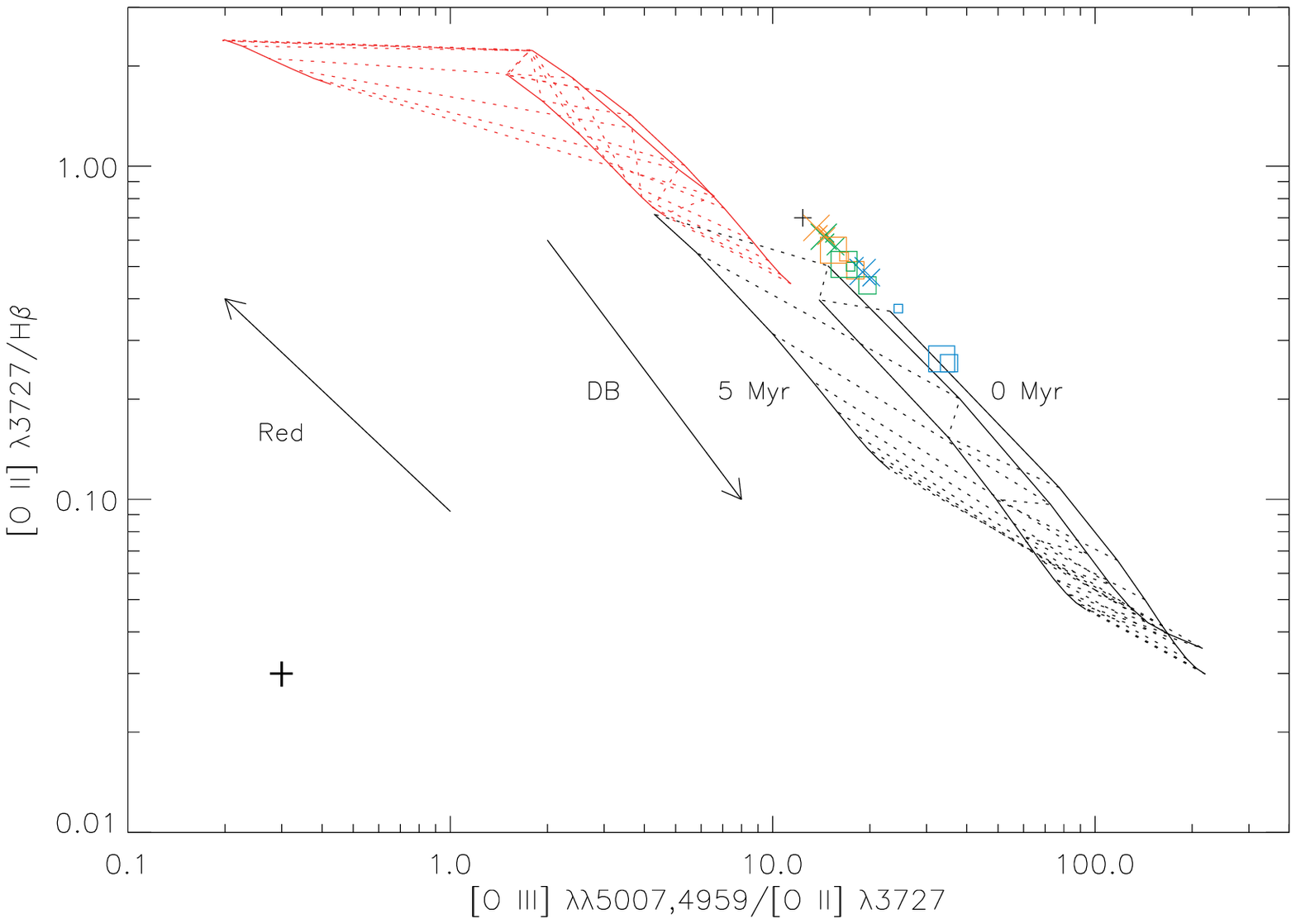}
\plotone{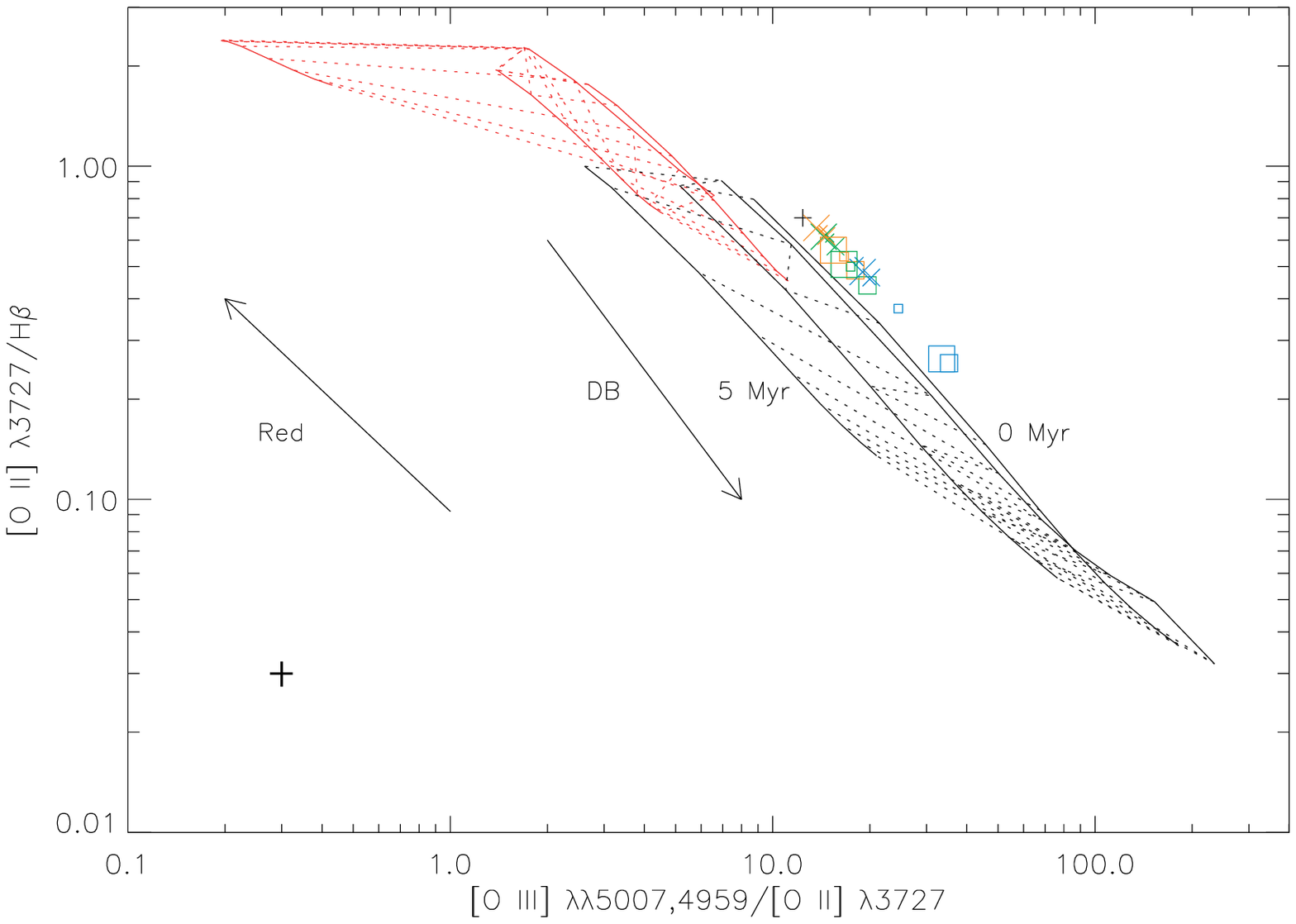}
\caption{The grid is described in Figure~\ref{fig_sb99n2}, and the symbols are described in Figure~\ref{fig_shock50}. Both panels show the CLOUDY and Starburst99 grids for $n_{\rm H}$=1000 cm$^{-3}$. The model in the top panel uses a filling factor of 0.1, and the bottom panel uses a filling factor of 0.01. The black cross indicates the observed ratios of Pea \peaBs.}
\label{fig_shock51}
\end{figure*}

Both shocks and photoionization may contribute to the observed nebular emission. To analyze the ionizing spectrum and the optical depth, therefore, we need to subtract the shock contribution. We use the \citet{allen08} Mappings III models at SMC and LMC metallicity and assume that all the observed \heii~$\lambda$4686 emission comes from shocks. We include emission from the shock's precursor, pre-shock gas that is photoionized by UV and X-ray emission produced in the shock \citep{allen08}. We then derive the intrinsic photoionized line ratios using
\begin{equation}
\label{eqn_shock1}
\frac{A}{H}=\frac{A_0-A_s}{H_0-H_s}=\frac{\frac{A_0}{H_0}X-\frac{A_s}{H_s}}{X-1},
\end{equation}where
\begin{equation}
\label{eqn_shock2}
X=\frac{\left (\frac{{\rm HeII}\lambda4686}{H}\right )_s}{\left(\frac{{\rm HeII}\lambda4686}{H} \right )_o}
\end{equation}and
\begin{equation}
\label{eqn_shock3}
{\rm HeII}\lambda4686_o={\rm HeII}\lambda4686_s.
\end{equation}In the above equations, $A$ is the emission line flux of interest, $H$ is the H$\beta$ line flux, the subscript $o$ indicates observed values, and the subscript $s$ indicates the modeled shock values. In addition to the two metallicities, we use the modeled shock values for magnetic field strengths of 0.5-10 $\mu$G and velocities from 300-1000 \kmps. Figures~\ref{fig_shock50} and~\ref{fig_shock51} compare the corrected photoionized line ratios for the two youngest Peas with the Starburst99 and CLOUDY models described in \S\ref{sec:nebular:wr}. 

The models show that when the additional flux contribution of shocks is taken into account, the optical depth implied by the nebular line ratios decreases. As the shock contribution to lower ionization lines, such as \oii, increases, the intrinsic \oiii/\oii~ratio increases, and the Peas move into the density-bounded, optically thin regime. Furthermore, the presence of shocks can raise the H$\alpha$/H$\beta$~ratio, leading to an overestimate of the dust extinction \citep{nakajima12}. If the dust abundance is overestimated, the Peas' \oii/H$\beta$~ratios should be lower, and their \oiii/\oii~ratios should be higher. This change would move the Peas farther into the optically thin regime, in a direction opposite to the reddening arrow in Figures~\ref{fig_sb99n2}-\ref{fig_sb99n3ff} and~\ref{fig_shock50}-\ref{fig_shock51}. The inferred optical depth of the Peas depends strongly on the inner nebular radius and shock properties assumed, however. In particular, lower velocity shocks have a larger effect on the inferred optical depth. The shock emission also depends on the density, which may differ from the density of the photoionized gas. The SMC and LMC shock models of \citet{allen08} have a fixed density of 1 cm$^{-3}$. The Mappings III solar metallicity models indicate that higher density shocks have less effect on the Peas' observed line ratios. If the density of the shock-heated gas is greater than 1 cm$^{-3}$, the inferred optical depth would be greater than that implied by Figures~\ref{fig_shock50}-\ref{fig_shock51}. An additional uncertainty comes from the modeled magnetic field strength, which is somewhat weaker than the Peas' magnetic fields. 

Although the presence of shocks shifts the Peas toward lower optical depths, the galaxies' line ratios still appear offset to higher \oii/H$\beta$~ratios and/or lower \oiii/\oii~ratios by factors of $\sim$1.5-2 relative to the $n<1000$ cm$^{-3}$ model grids (see e.g., Figure~\ref{fig_shock50}). One possible explanation for this offset is that higher density models are more appropriate for the extreme Peas. If shocks are indeed present, they may contribute a substantial fraction of the observed \sii~emission, which may lead to incorrect density estimates (\S\ref{sec:data}). The existence of lower ionization nebulae within the Peas may also resolve the discrepancy. If low ionization nebulae contribute a fraction of the observed \oii~emission, they could account for the higher \oii/H$\beta$~and lower \oiii/\oii~ratios relative to the grids. Adopting lower reddening corrections or higher metallicities for the Peas could also lessen the offset; however, the uncertainties in these quantities are too low to completely explain the disagreement. Finally, as mentioned in \S\ref{sec:nebular}, a population of hot, low-metallicity stars could provide an additional means of increasing the \oiii/\oii~ratio. Given the current lack of evidence for such stars, however, we conclude that the presence of multiple nebulae and higher densities in the star-forming regions are the more likely explanations.

Despite these uncertainties, the CLOUDY models show that including shock emission lowers the inferred optical depths of the Peas, in some cases dramatically. If the shocks are due to SN activity, the Peas must be old enough for such events to take place. As a test, we consider Pea \peaCs, the highest EW Pea and potentially the youngest. Our results for the other extreme Peas are comparable. From the observed \heii~and the modeled \heii/H$\alpha$ ratios, we estimate how many SNe are necessary to account for the \heii~emission. Assuming all the \heii~is due to shocks, the modeled \heii/H$\alpha$~ratios imply that shocks provide $5.6 \times 10^{40}$-$4.5 \times 10^{41}$\ergps~of Pea \peaCs's H$\alpha$ luminosity. The H$\alpha$~luminosities of SNRs range from $\sim10^{36} $ to $\sim10^{38}$ \ergps \citep{degrijs00}. If we assume an intermediate SNR luminosity of 10$^{37}$ \ergps, then Pea \peaCs~would need 5600-45,000 SNe ($\sim$4-50\% of its O stars) to reach the necessary \heii~and H$\alpha$~luminosities. While the former value could be reached in less than 4 Myr following an instantaneous burst, the latter value requires more than 6 Myr. However, if the SNRs are extremely bright (10$^{38}$\ergps), as might be expected for a young burst, than the required number of SNe drops to only 560-4500, 0.4-5\% of Pea \peaCs's O stars. The first SNe at $\sim$3.5-4 Myr could provide the necessary H$\alpha$ and \heii~luminosities. The ages for the continuous star formation case are similar; 5000 SNe would go off within the first 5 Myr. 

One example of a Pea-like galaxy that may have shock-induced \heii~emission is II Zw 40, a nearby BCD that shares many properties with the Green Peas. It has a metallicity of 0.2\Zsol and an H$\beta$~EW of 272\AA~\citep{guseva00}, suggesting a similar age to the Peas. In addition, II Zw 40 has detectable WR features, including the blue bump and \siiii~$\lambda$4565, but the blue bump is dominated by the much stronger nebular emission of \heii, \feiii, and \ariv~\citep{guseva00}. The much weaker strength of the WR features compared to the nebular emission shows that these features could easily be below our detection limits. Based on the similarities between \heii-emitting galaxies with and without WR features, \citet{guseva00} suggest that shocks generate some of the nebular \heii~emission, even in WR galaxies. The presence of weak \fev~$\lambda$4227 emission and high-velocity emission-line wings in II Zw 40 supports the shock hypothesis \citep{thuan05}. Although II Zw 40 is unlikely to be optically thin \citep[e.g.,][]{joy88,beck02}, it shows that shock ionization of \heii~is feasible at the Peas' metallicities and ages. Thus, even if WR stars are present in the Green Peas, shock heating may still be the primary origin of the high ionization nebular lines. 

The above analysis demonstrates that shock-heating from the earliest SNe could account for the observed \heii~emission. In addition, shocks due to a galaxy merger or gas infall could occur even earlier, and younger ages for the Peas could be possible. If shocks are present, the Peas' high \oiii/\oii~ratios may indicate a low optical depth.

\section{Consequences for the Optical Depth}
\label{sec:discussion}

The source of the high ionization emission in the extreme Green Peas could be WN9 stars, early-type WR stars, chemically homogeneous O stars, shocks, or some combination of these sources. Each of these sources has different implications for the escape of ionizing radiation, which we now discuss. 

If the \heii~emission is a stellar feature or due to stellar photoionization, the Peas' optical depth is ambiguous. Low density models, with $n < 100$ cm$^{-3}$ indicate a high optical depth. At higher densities, consistent with the Peas' \sii~ratios, whether or not the Peas are optically thin depends on the inner nebular radius and the filling factor. For instance, decreasing the filling factor of the dense gas decreases the inferred optical depth (see Figures~\ref{fig_sb99n2} and \ref{fig_sb99ff}). 

Increasing the inner radius lowers the ionization parameter; the high \oiii/\oii~ratios we observe would then have to be due to a low optical depth. Unlike the CLOUDY models, the emission from the Peas is not likely to originate from a single \hii~region. However, if the emission is dominated by one cluster or several similar clusters, we can estimate which CLOUDY model geometries may be most reasonable. 

The stellar winds of WR stars may blow cavities of 50 \pc~or more in the ISM \citep[e.g.,][]{oey96}, and SN explosions will expand the superbubble cavity further. The broad wings in the Peas' H$\alpha$ emission may already indicate that stellar feedback is evacuating gas around the star clusters and driving galactic winds \citep{amorin12b}. \citet{strickland99} simulate a superbubble around a 10$^5$\Msol~cluster at 0.25\Zsol, expanding into an ambient medium with density $n = 100$ cm$^{-3}$. The superbubble cavity reaches a radius of 50 \pc~within 3.5 Myr and expands to 70 \pc~within 4.5 Myr. The radius scales with cluster mass as $R\propto (M/n)^{1/5}$ \citep{castor75,strickland99}, so for our 10$^6$\Msol~Starburst99 simulation, the cavity radii should be 79 and 111 \pc, respectively. If we increase the density by a factor of 10, as appropriate for Pea \peaBs, the radii remain 50-70 \pc. Therefore, for our $10^6$\Msol~starburst CLOUDY models, the models with inner radii greater than 50 pc may have the most realistic geometries. Models with these inner radii and $n > 100$ cm$^{-3}$ generally indicate a low optical depth (see Figure~\ref{fig_sb99n3ff}). 

The exact optical depth of the Peas also depends on their age and the resulting ionizing spectrum. For instance, Figure~\ref{fig_sb99n2} shows that for a given inner radius, a 5 Myr-old model gives a slightly lower optical depth than a 4 Myr-old model. The ionizing spectrum of chemically homogeneous O stars is highly uncertain, and the existence of 60,000 K homogeneous stars has not been observationally confirmed. We cannot determine the optical depth of the Peas if their ionization is due to a chemically homogeneous population. Thus, while we cannot assess the precise optical depth of the extreme Peas, a low optical depth remains plausible, particularly for the Peas with the highest densities.

As discussed in \S\ref{sec:shocks}, including a shock contribution to the Peas' line fluxes decreases estimates of their optical depth. This same situation is apparent in the spectra of DEM L301, an LMC superbubble. \citet{oey00} show that a combination of shock-heating and density-bounding explain DEM L301's emission, although at first glance, the line ratios appear to indicate radiation-bounding. \citet{voges08} confirm that DEM L301 is indeed optically thin. 

Not only do SNe affect the nebular emission, but they may actually promote a low optical depth. As SN-driven bubbles expand in the ISM and eventually blow out of the galaxy, they create low-density passages that aid photon escape \citep{clarke02}. Like Lyman-break galaxies, the Peas have large SFRs and spatially distributed star formation, which enhance the ISM porosity \citep{clarke02}. In addition, with their young ages, the Peas may be at the optimal time for the escape of Lyman-continuum radiation; they are old enough for SNe and stellar winds to begin to reshape the ISM, but young enough to possess large numbers of UV-luminous massive O or WR stars.

The CLOUDY models for the WR and shock-heating scenarios at $n=100$ cm$^{-3}$ are generally optically thick or borderline  optically thin. Models at higher densities favor lower optical depths but could still be consistent with optically thin nebulae if the inner radii are sufficiently small and shocks are negligible. Nevertheless, at all densities, the Peas may be more optically thin than the models suggest. For instance, optically thick clumps tend to dilute the signature of density-bounding \citep[e.g.,][]{oey00,giammanco05}. The CLOUDY models further assume that the emission originates from a single nebular region or identical nebulae. In reality, the emission is the sum of the emission from multiple nebulae, which may have different levels of ionization. Low-ionization emission from other nebulae or a companion galaxy could lower the observed \oiii/\oii~ratios and cause the gas to appear more optically thick than it actually is. 

The extreme Peas may therefore have high escape fractions for ionizing radiation. However, further observations are needed to diagnose their optical depths and distinguish among the potential ionizing sources. We are obtaining deeper spectra of the extreme Peas to confirm or refute the existence of shocks and to constrain the WR population. Observations of massive stars in low-metallicity galaxies are also critical to understand homogeneous stellar evolution and the stellar spectra of low-metallicity O and WR stars.

\section{Summary}
\label{sec:summary}

The Green Peas are a rare class of compact, emission-line galaxies at $z=0.1-0.3$ with unusually powerful \oiii~emission and the potential for a high escape fraction. Through their high ionization emission, extreme star-formation intensities, and clumpy morphologies, the Green Peas provide a glimpse of massive star feedback and ionizing photon propagation in high redshift galaxies. While the large \oiii/\oii~ratios of the Peas may indicate a low optical depth, high ionization lines such as \heii~and \ariv~in the SDSS spectra show evidence for strong, high-energy emission. Using stellar population and nebular photoionization models, we have investigated the ionizing sources and optical depths for six of the highest ionization Peas. 

The six extreme Peas are young and powerful starbursts, with high electron temperatures of $\sim 15,000$ K and densities of 100-1000 cm$^{-3}$. Their high Balmer-line EWs indicate young ages of 3.7-5.1 Myr, according to Starburst99 instantaneous burst models. We detect the \hei~$\lambda$3819 line in emission in several objects and in stacked spectra, which corroborates the young ages. The $\lambda$3819 line sets an upper limit of 3 Myr on the ages for an instantaneous burst and 10 Myr for continuous star formation \citep{gonzalez99}.

The Peas' young ages and the presence of \heii~$\lambda$4686 emission set strong constraints on the possible ionizing sources in the Peas. No WR features are visible in stacked spectra of the Peas. However, we tentatively detect WR features in Peas \peaCs, \peaDs, and \peaFs. The narrow FWHMs (3-5 \AA) of the \heii~emission suggest a nebular origin or a narrow WNL stellar feature.

Using the WNL line luminosities of \citet{brinchmann08} and \citet{guseva00}, we calculate the number of WNL stars necessary to produce the observed $\lambda$4686 emission as a stellar feature. The high resulting WR/O ratios and the $3\sigma$ \siiii~$\lambda$4565 detection limits rule out the required WNL population for the $Z < 0.2$\Zsol~case, but a WNL population with $Z \geq 0.2$\Zsol~is possible. However, a higher-metallicity WNL population may not be consistent with the Peas' low measured metallicities.

Alternatively, the observed \heii~may result from photoionization by hot stars. CLOUDY photoionization models \citep{ferland98} of single-star \hii~regions demonstrate that ordinary O stars are not hot enough to provide the necessary He$^+$-ionizing flux. CLOUDY models using early-type WR stars, with core temperatures above 70,000 K, fit most of the observed line ratios. From the ionizing photon rates of \citet{smith02}, we estimate the number of WNE and WCE stars needed to produce the observed \heii. The resulting WR/O ratios are consistent with Starburst99 instantaneous burst models, but the lack of broad \heii~emission in most objects eliminates the WCE stars as the dominant cause of the emission. Starburst99 and CLOUDY models likewise show that only 4-5 Myr starbursts, which contain a substantial WR population, can simultaneously match the Peas' \oiii/\oii~and \heii~emission. 

Putative chemically homogeneous O stars could explain \heii~emission in galaxies without WR features \citep{shirazi12}. These stars may form at low metallicity \citep[e.g.,][]{maeder87,yoon06} and can theoretically reach effective temperatures much higher than ordinary O stars \citep{brott11}. CLOUDY single-star \hii~region models with the 60,000 K stellar model of \citet{kudritzki02} demonstrate that such hot O stars could account for the observed \heii. In addition, the presence of extremely hot O stars may be necessary to match the observed \oii/H$\beta$ and \oiii/\oii~ratios of the Peas. More research is needed to determine whether homogeneous O stars exist, however, and how many ionizing photons they produce.

We rule out an AGN contribution to the \heii~emission at the 10\% level, using the diagnostic diagram of \citet{shirazi12}.
Likewise, we find that HMXBs cannot be the source of the \heii~ionization. Using the models of \citet{linden10}, we calculate the expected number of luminous HMXBs in the Peas. Even assuming that each HMXB is a ULX, with the high nebular \heii~$\lambda$4686 luminosity of Holmberg II X-1 \citep{kaaret04}, the resulting emission is an order of magnitude too low.

Fast, radiative shocks are an alternative mechanism for the production of \heii~\citep{guseva00,thuan05}, and the existence of fast shocks is supported by the high velocity Balmer-line wings \citep{amorin12b} and \oi~and \none~emission in the Peas. To account for the contribution of shocks to the observed emission, we use the line emission predictions of the Mappings III shock models \citep{allen08} for different magnetic field strengths and velocities. Assuming shocks generate all of the \heii~emission, we subtract the shock contribution from the observed line ratios. After correcting for shock emission, the nebular line ratios indicate a lower optical depth, particularly for slower shocks. Based on the H$\alpha$~luminosities from the shock models, the first SNe, exploding 3.5-4 Myr after an instantaneous burst, could produce the observed \heii~emission. Younger ages could be possible if infalling gas or mergers are the cause of the shocks.

Although the precise origin of the \heii~emission is still uncertain, the CLOUDY models show that the Peas may be optically thin. Models with higher densities and larger inner radii imply lower optical depths. If the \heii~is due to WR stars, models with inner radii $\geq$ 50 \pc~and densities $\sim$ 100 cm$^{-3}$ support both optically thick and optically thin scenarios, while models with the same inner radii and  $n=1000$ cm$^{-3}$ support a low optical depth. Including the effects of shocks lowers the estimated optical depth further, and the presence of density inhomogeneities or emission from multiple nebulae may also lead us to overestimate the optical depth.

The low levels of extinction, intense star formation, and young ages of the Green Peas make them ideal candidates for escaping ionizing radiation. The nebular emission of the Peas with the highest \oiii/\oii~ratios, their high densities, and the likely presence of shocks in these objects suggest that they may be leaking ionizing photons into the IGM, and their ages may optimize photon escape. In these 3-4 Myr starbursts, stellar feedback is likely beginning to carve out passages in the ISM for the escape of ionizing photons, while large numbers of massive stars are still available to produce Lyman-continuum photons. However, the data are still inconclusive, and future observations are necessary to establish the escape fraction of the Green Peas. These galaxies offer a rare opportunity to study high-redshift star formation conditions, the effects of stellar feedback on the ISM, and the propagation of ionizing photons from star clusters to the IGM.  

\acknowledgments{We thank the anonymous referee for constructive feedback and Eric Pellegrini and Jordan Zastrow for their assistance with CLOUDY. A.E.J. acknowledges support from an NSF Graduate Research Fellowship, and M.S.O. acknowledges support from NSF grant AST-0806476. }

\end{document}